\documentclass[useAMS,usenatbib,letter]{mn2e}

\usepackage{graphicx,url,color,rotating,amssymb,amsmath,hyperref}
\everymath{\displaystyle}
\usepackage{dcolumn}
\usepackage{bm}
\usepackage{sansmath}

\newcommand{\beq}{\begin{equation}}
\newcommand{\eeq}{\end{equation}}
\newcommand{\bal}{\begin{aligned}}
\newcommand{\eal}{\end{aligned}}
\newcommand{\beqa}{\begin{eqnarray}}
\newcommand{\eeqa}{\end{eqnarray}}
\newcommand{\avg}[1]{\left\langle{#1}\right\rangle}
\newcommand{\erf}{{\rm erf}}

\newcommand{\hiMsun}{h^{-1} \rm M_\odot}
\newcommand{\hiMpc}{h^{-1} \rm Mpc}
\newcommand{\hikpc}{h^{-1} \rm kpc}
\newcommand{\hoverMpc}{h\,{\rm Mpc^{-1}}}

\newcommand{\tdelta}{\tilde{\delta}_m}
\newcommand{\tu}{\tilde{u}}
\newcommand{\tug}{{\tilde{u}}}
\newcommand{\rhog}{{n_{\rm gal}}}
\newcommand{\brhog}{{{\bar n}_{\rm gal}}}
\newcommand{\deltag}{{\delta_{\rm gal}}}
\newcommand{\tdeltag}{{\tilde\delta_{\rm gal}}}
\newcommand{\tdeltagz}{{\tilde\delta_{\rm gal}^z}}
\newcommand{\deltaD}{\delta_{D}}
\newcommand{\vx}{{\bm{x}}}
\newcommand{\vecr}{{\bm{r}}}
\newcommand{\vk}{{\bm{k}}}
\newcommand{\Mvir}{M_{\rm vir}}
\newcommand{\Rvir}{R_{\rm vir}}
\newcommand{\cvir}{c_{\rm vir}}
\newcommand{\Nc}{N_{\rm cen}}
\newcommand{\Ns}{N_{\rm sat}}
\newcommand{\kmax}{k_{\rm max}}
\newcommand{\fsys}{f_{\rm sys}}
\newcommand{\fsky}{f_{\rm sky}}
\newcommand{\sigmav}{\sigma_{\rm v}}
\newcommand{\bv}{b_{\rm v}}

\begin{document}
\title[Theoretical systematics in galaxy clustering]
{\LARGE The impact of systematic uncertainties in $\bm N$-body simulations
on the precision cosmology from galaxy clustering: A halo model approach}
\author[H.-Y.\,Wu and D.\,Huterer]{Hao-Yi Wu\thanks{E-mail: hywu@umich.edu} and
Dragan Huterer\thanks{E-mail: huterer@umich.edu} \\ Department of
Physics, University of Michigan, 450 Church St, Ann Arbor, MI
48109-1040, USA}
\date{Accepted 2013 June 26. Received 2013 May 30; in original form 2013 March 4}

\maketitle

\begin{abstract}
Dark matter $N$-body simulations provide a powerful tool to model the
clustering of galaxies and help interpret the results of galaxy
redshift surveys. However, the galaxy properties predicted from
$N$-body simulations are not necessarily representative of the
observed galaxy populations; for example, theoretical uncertainties
arise from the absence of baryons in $N$-body simulations.  In this
work, we assess how the uncertainties in $N$-body simulations impact
the cosmological parameters inferred from galaxy redshift surveys.
Applying the halo model framework, we find that the velocity bias of
galaxies in modelling the redshift-space distortions is likely to be
the predominant source of systematic bias.  For a deep, wide survey
like BigBOSS, current 10 per cent uncertainties in the velocity bias
limit $\kmax$ to 0.14 $\hoverMpc$.  In contrast, we find that the
uncertainties related to the density profiles and the galaxy
occupation statistics lead to relatively insignificant systematic
biases.  Therefore, the ability to calibrate the velocity bias
accurately -- from observations as well as simulations -- will likely
set the ultimate limit on the smallest length scale that can be used
to infer cosmological information from galaxy clustering.
\end{abstract}


\section{Introduction}

The large-scale distribution of galaxies has been used to probe the
structure and composition of the universe for over three decades.
From the pioneering analyses of the Lick catalogue
\citep{Groth_Peebles_97} and the CfA Redshift Survey
\citep{Huchra83,GellerHuchra89} revealing the cosmic web, the APM
Galaxy Survey hinting the departure from the standard cold dark matter
model \citep{Maddox90} to the subsequent 2dF Galaxy Redshift Survey
\citep{Colless01}, the Sloan Digital Sky Survey \citep[SDSS;][]{SDSS}
and the VIMOS-VLT Deep Survey \citep{LeFevre05}, galaxy redshift
surveys have revolutionized the view of the large-scale structure of
the universe.  Recently, the WiggleZ Dark Energy Survey
\citep{Drinkwater10} and the SDSS-III Baryon Oscillation Spectroscopic
Survey (BOSS; \citealt{Schlegel09}) have measured the galaxy
clustering to unprecedented precision and provided stringent
constraints on the cosmological parameters.

One of the most important features in the galaxy clustering is the
baryon acoustic oscillations (BAO), originating from the waves in the
primordial electron--photon plasma before the recombination.  The
sound horizon at the end of recombination is manifested as a peak in
the real-space two-point correlation function or as wiggles in the
Fourier-space power spectrum.  This characteristic scale of BAO is
considered as a standard ruler of the different evolution stages of
the universe, and as a dark energy probe with relatively
well-controlled systematics \citep{BlakeGlazebrook03,Seo03}.  Indeed,
since its discovery \citep{Miller01,Cole05,Eisenstein05}, BAO has been
providing ever improving constraints on cosmological parameters
\citep[e.g.,][]{Percival10,Blake11b,Anderson12}.

Beyond the BAO feature, the full scale-dependence of the clustering of
galaxies contains much more information and can be used to constrain
cosmology \citep[e.g.,][]{Tegmark06,Reid10,Tinker12,Cacciato12} and
the halo occupation statistics
\citep[e.g.,][]{Abazajian05,Tinker05,vdBosch07,ZhengWeinberg07,Zehavi11}.
From the perspective of power spectrum $P(k)$, the number of modes
increases as $k^3$, and the information content increases dramatically
as one goes to smaller scales.  However, when one tries to draw
information from high $k$, especially at low redshift, the density
perturbations become non-linear and difficult to model
\citep[e.g.,][]{Smith03,Heitmann10,Jennings11}, which can introduce
significant systematic errors in the recovered cosmological parameters
\citep[e.g.,][]{dlTorre12,Smith12}.

The analysis of galaxy clustering often relies on $N$-body simulations
and synthetic galaxy catalogues to model the non-linearity on small
scales, as well as to estimate the cosmic and sample covariances.  For
example, the WiggleZ team has validated their model for the non-linear
galaxy power spectrum using the GiggleZ Simulation\footnote{\tt
http://tao.it.swin.edu.au/partner$\mbox{-}$resources/\\simulations/gigglez/}
\citep{Parkinson12}, while synthetic galaxy catalogues based on the
Large Suite of Dark Matter Simulations (LasDamas\footnote{\tt
http://lss.phy.vanderbilt.edu/lasdamas/}) have been used in the galaxy
clustering analysis of SDSS \citep{ChuangWang12, Xu13}.

For upcoming surveys, synthetic catalogues generated from $N$-body
simulations will likely be routinely used to calibrate galaxy surveys.
However, $N$-body simulations are not free from systematics.  In
$N$-body simulations, galaxies are assigned to haloes or dark matter
particles based on models such as halo occupation distribution
\citep[HOD;][]{PeacockSmith00,Scoccimarro01,BerlindWeinberg02},
abundance matching \citep{Kravtsov04,ValeOstriker04}, or semi-analytic
models \citep{WhiteFrenk91,Kauffmann93,SomervillePrimack99,Cole00}.
The galaxy populations predicted by simulations can be affected by
intensive stripping in dense environment
\citep[e.g.,][]{WetzelWhite10} and the absence of baryons
\citep[e.g.,][]{Weinberg08,Simha12}.  On the other hand, when one uses
dark matter particles to model the behaviour of galaxies, systematic
errors may arise because the positions and velocities of galaxies do
not necessarily follow those of dark matter particles
(\citeauthor{Wu12b} 2013a).  Hydrodynamical simulations that include
proper treatments of baryonic physics can be another avenue to predict
the properties of galaxies more reliably; however, because these
simulations are more computationally intensive, it is not yet
practical to use them to achieve the statistics and high resolution
required by upcoming large surveys.

In addition, it has been shown that galaxies predicted from $N$-body
simulations cannot recover the spatial distribution of observed
galaxies.  For example, Wu et al.~(in preparation) have shown that in
high-resolution $N$-body simulations of galaxy clusters, subhaloes
tend to be prematurely destroyed and fail to predict the location of
galaxies (also see Appendix \ref{app:Consuelo}). The need to include
``orphan galaxies'' (galaxies not associated with subhaloes in
simulations) to improve the completeness of predicted galaxies has
been frequently addressed in the community
\cite[e.g.,][]{Gao04,Wang06,Guo11}; however, even including orphan
galaxies does not lead to consistent galaxy clustering at all scales .
For example, \citet{Guo11} have shown that the galaxy population
generated using the semi-analytic model applied to the Millennium
Simulations overestimates the small scale clustering (also see
\citealt{Contreras13}).

In this paper, we examine the impact of the systematics in $N$-body
simulations on the predictions of galaxy clustering.  We calculate the
galaxy power spectrum based on the halo model, with inputs from the
results of recent $N$-body simulations.  We use the information of the
full power spectrum of galaxies to forecast the cosmological parameter
constraints and determine at which scale these systematics start to
become relevant.  We specifically explore how these uncertainties will
limit our ability to utilize the cosmological information from small
scale.

This paper is organized as follows.  In Section \ref{sec:review}, we
review the halo model prediction for galaxy power spectrum.  In
Section \ref{sec:fiducial}, we present our fiducial assumptions and
discuss the information content associated with $P(k)$.  Section
\ref{sec:HOD} explores the self-calibration of HOD parameters.
Section \ref{sec:MF} addresses the impact of the uncertainties in the
halo mass function on the cosmological constraints from galaxy
clustering.  Section \ref{sec:sys} focuses on various systematics
associated with the properties of galaxies in dark matter haloes in
$N$-body simulations and presents the required control of these
sources of systematic error.  We conclude in Section
\ref{sec:summary}.  In Appendix \ref{app:Consuelo}, we present the
galaxy number density profile model used in this work.  In Appendix
\ref{app:halomodel}, we provide detailed derivation of the galaxy
power spectrum based on the halo model. In Appendix~\ref{app:Cij}, we
derive the power spectrum covariance.

\section{Halo model and galaxy power spectrum: a review}\label{sec:review}

Throughout this work, we use the power spectrum of galaxies $P(k)$ as
our clustering statistic.  Possible alternatives include the
three-dimensional correlation function $\xi(r)$ and its
two-dimensional analogue -- the angular two-point function $w(\theta)$
or the projected two-point function $w_p(r_p)$.  While the
Fourier-space power is more difficult to measure from the galaxy
distribution, it is `closest to theory' in the sense that the other
aforementioned quantities are weighted integrals over
$P(k)$. Therefore, it is easiest to see the effect of the
uncertainties in theoretical modelling by using the power spectrum.
While these different functions measured in a given galaxy survey
contain the same information in principle, in data analysis sometimes
discrepancies occur \citep[e.g.,][]{Anderson12}.

\subsection{Basic model}\label{sec:Pk_basic}

In this section, we provide the key equations of the galaxy power
spectrum derived from the halo model, following
\cite{ScherrerBertschinger91}, \cite{Seljak00}, and
\cite{CooraySheth02}.  The detailed derivation is provided in Appendix
\ref{app:halomodel}.

The halo model assumes that all galaxies are inside dark matter
haloes.  To model the distribution of galaxies, we need the following distributions.
\begin{enumerate}
\item Statistics and spatial distribution of dark matter haloes:
  \begin{itemize}
 	 \item Halo mass function, ${\rm d}n/{\rm d}M$, the number density of haloes as a
 	   function of the halo mass.
 	 \item Halo bias, $b^2(M) = P_{\rm hh}(k)/P_{\rm lin}(k)$, where $P_{\rm hh}$ is the
	    power spectrum of haloes and $P_{\rm lin}$ is the linear matter power
  	  spectrum.  We limit our use of $b(M)$ to large scales where $b(M)$ is
   	 scale independent.
  \end{itemize}
\item Statistics and spatial distribution of galaxies in a halo:
  \begin{itemize}
	  \item HOD function,  $P(N|M)$, the
  	  probability distribution function of the number of galaxies in a
  	  halo of a given mass.  The number of galaxies $N$ is further split into 
   	 the contribution from central galaxies $\Nc$ (0 or 1) and from satellite galaxies $\Ns$. 
  	 \item Galaxy number density profile, $u(r|M)$, the radial
 	    dependence of the galaxy number density inside a halo of a given mass.
  	    We normalize $u$ such that
   	  $\int u(r|M){\rm d}^3\vecr = 1$.
   	  We  also use the density profile in Fourier space,
	$\tug(\vk|M) = \int {\rm d}^3\vx u(\vx|M) {\rm e}^{-i\vk\cdot\vx} \ , $
	and $\tug \rightarrow 1$ for small $k$.
  \end{itemize}
\end{enumerate}
The mean galaxy number density is given by 
\beq
{\brhog}\equiv \avg{\rhog}= \int dM \frac{dn}{dM} \avg{N|M}  \ .
\eeq
The power spectrum is contributed by two galaxies in two different
haloes (the two-halo term, $P_{\rm gg}^{\rm 2h}$)
and two galaxies in the same halo (the one-halo term, $P_{\rm gg}^{\rm 1h}$):
\beqa
P(k) &=& P_{\rm gg}^{\rm 1h}(k) + P_{\rm gg}^{\rm 2h}(k) \\[0.1cm]
P_{\rm gg}^{\rm 2h}(k) &=& \left[\frac{1}{{\brhog}} \int dM \frac{dn}{dM} \avg{N|M} b(M) \tug(k|M)
\right]^2 P_{\rm lin}(k)  \nonumber\\[-0.2cm]\\
P_{\rm gg}^{\rm 1h}(k)&=&\frac{1}{{\brhog}^2} \int dM \frac{dn}{dM}
\avg{\displaystyle{\left . N\choose 2\right |}M} f(k|M)
\eeqa
Here $\avg{X|M}$ indicates the average value of quantity $X$
at a given halo mass $M$.
In the one-halo term,
\beq\bal
\avg{\displaystyle {\left . N\choose2\right |}M} 
f(k|M)  
&= \left[
\avg{\Ns|M}\tug(k|M) \right . \\[-0.1cm]
&+ \left . \frac{1}{2} \avg{\Ns(\Ns-1)|M}  |\tug(k|M)|^2  
\right] \, , \nonumber
\eal\eeq
which takes into account the contribution from central--satellite and satellite--satellite pairs \citep{BerlindWeinberg02}.

\begin{figure*}
\includegraphics[width=\columnwidth]{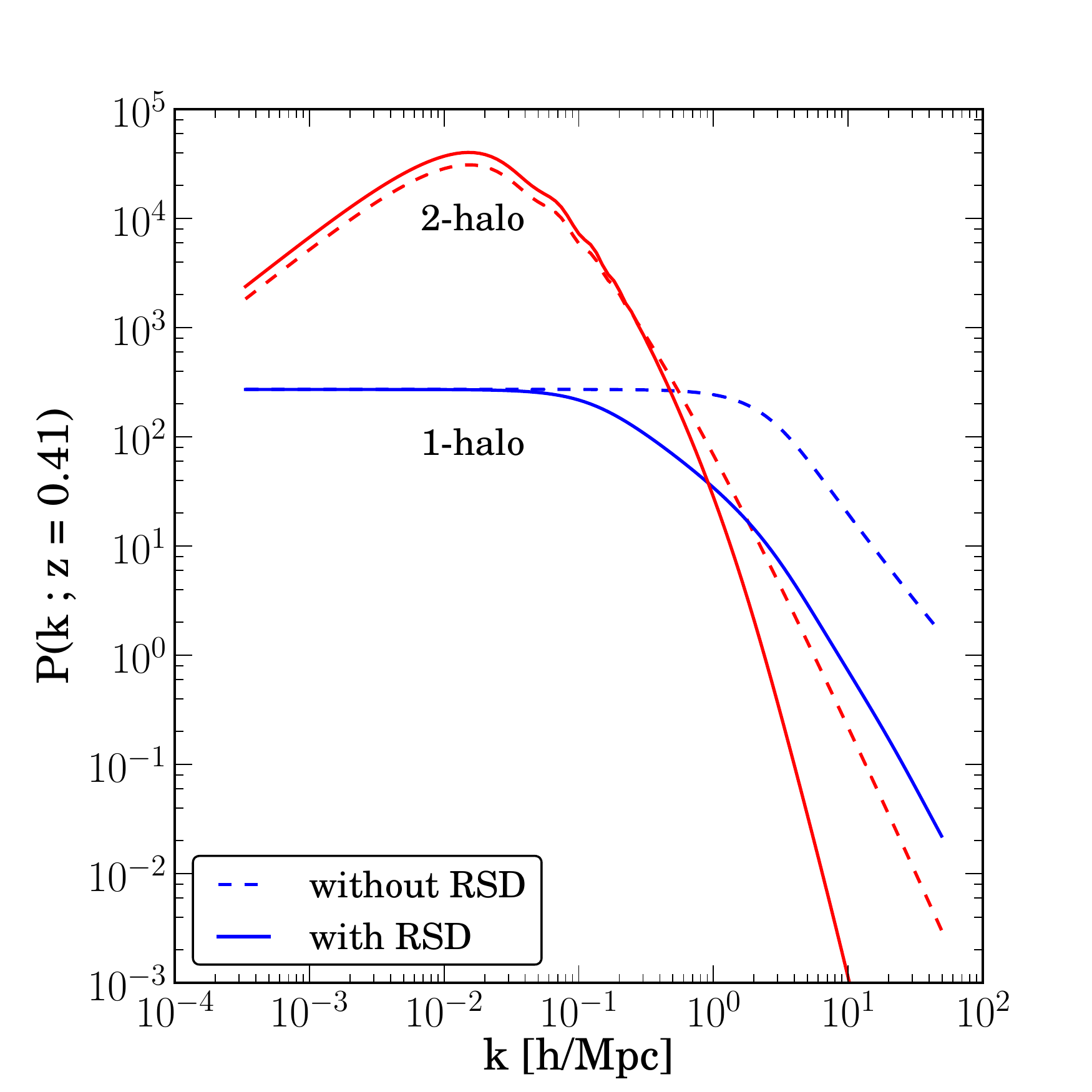}
\includegraphics[width=\columnwidth]{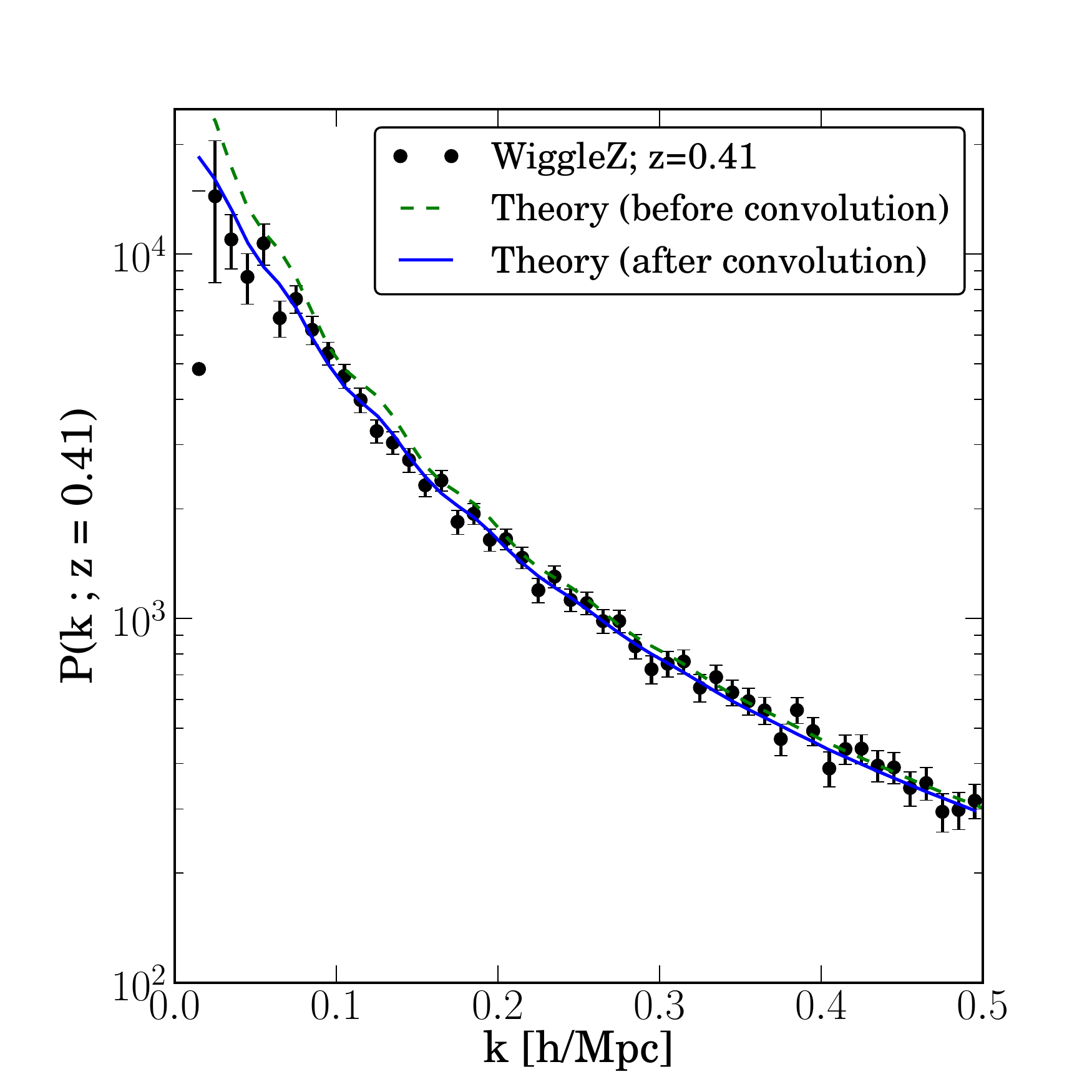}
\caption{Galaxy power spectrum calculated based on the halo model.
Left: the blue and red curves show the one- and two-halo terms,
respectively. The solid curves include the RSD, while the dashed
curves do not.  RSD greatly reduce the power spectrum at small scale
(the `Fingers-of-God' effect), and only slightly shift the scale where
one- and two-halo terms cross.  Right: Our model fit to WiggleZ data
from \citet{Parkinson12}.  The blue solid curve shows the theoretical
$P(k)$ with the best-fitting HOD parameters and has been convolved
with the observational window function.  }
\label{fig:PS_rsd}
\end{figure*}

\subsection{Redshift-space distortions}\label{sec:Pk_rsd}

In observations, one cannot recover the exact three-dimensional
spatial distribution of galaxies, because the redshifts of galaxies
are impacted by their motions due to the local gravitational field and
do not reflect their true distances.  On larger scales, galaxies tend
to move towards high-density regions along filaments, and these
motions tend to squash the galaxy distribution along the line of sight
and boost the clustering, a phenomenon known as the Kaiser effect
\citep{Kaiser87}.  On small scales, the virial motions of galaxies
inside a halo tend to make the galaxy distribution in the redshift
space elongated along the line-of-sight, causing the so-called
Fingers-of-God effect and reducing the small-scale power.  In this
section, we briefly describe the model we use for the redshift-space
distortions (RSD) for $P(k)$, following \cite{Seljak01},
\cite{WhiteM01}, and \cite{CooraySheth02}.  We adopt one of the
simplified models -- assuming the velocity distribution function to be
Gaussian -- and note that the improvement of the RSD model is
currently an active research area.

Since the one-halo term involves the halo scale, we only consider the
virial motions of galaxies inside a halo, which can be modelled as
\citep{Peacock99}
\beq
\tdeltagz (\vk) = \tdeltag(\vk) {\rm e}^{-\frac{1}{2}[k\sigmav(M)\mu]^2} \ ,
\eeq
where 
$\tdeltagz$ and $\tdeltag$ are the number density fluctuations of galaxies
with and without the effect of RSD,
$\sigmav(M)$ is the velocity dispersion of galaxies 
inside a halo
of mass $M$ and $\mu= \hat{\vk}\cdot \hat{\vecr}$. 
We average over $\mu$ to obtain the angular averaged one-halo term
\beq
P_{\rm gg}^{\rm 1h}(k) = \frac{1}{{\brhog}^2} \int dM \frac{dn}{dM} 
\avg{\displaystyle{\left . N\choose2\right |} M}
\,f_R(k|M) \ ,
\label{eq:P1h_rsd}
\eeq
where
\begin{equation}
\begin{aligned}
\avg{\displaystyle{\left . N\choose2\right | } M} 
&f_R(k|M)
=\left[
\avg{\Ns|M}\tug(k|M) R_1(M) \right .\\
&+ \left . \frac{1}{2} \avg{\Ns(\Ns-1)|M}  |\tug(k|M)|^2  R_2(M)
\right]  \ .
\nonumber
\end{aligned}
\end{equation}
The factor
\beq
R_p(M) = \frac{\sqrt{{\rm{\rm\pi}}}}{2}\frac{\erf[k\sigmav(M)\sqrt{p/2}]}{k\sigmav(M)\sqrt{p/2}}
\eeq
comes from averaging over $\mu$.

For the two-halo term, we multiply the large-scale and small-scale
effects together (see \citealt{Peacock99} and section 4 in
\citealt{PeacockDodds94})
\beq
\tdeltagz (\vk) = \left(\tdeltag(\vk) + f(\Omega_{\rm M}) \tdelta(\vk) \mu^2 \right) {\rm e}^{-\frac{1}{2}[k\sigmav(M)\mu]^2} \ .
\eeq
The first part is the familiar Kaiser result with $f(\Omega_{\rm
M})\equiv d\ln D/d\ln a$, where $D(a)$ is the linear growth function
of density fluctuations and $a$ is the scale factor.  The density
fluctuation of dark matter is denoted by $\tdelta$.  The calculation
thus includes not only the galaxy power spectrum, but also the matter
power spectrum and the matter--galaxy cross power spectrum.  After
averaging over $\mu$, we obtain
\beq
P_{\rm gg}^{\rm 2h}(k)= \left(F_g^2 + \frac{2}{3}F_g F_v + \frac{1}{5} F_v^2 \right) P_{\rm lin}(k) \ ,
\label{eq:P2h_rsd}
\eeq
where 
\beq
F_g(k) = \frac{1}{{\brhog}} \int dM \frac{dn}{dM} \avg{N|M}  b(M) R_1(M) {\tu(k|M)}
\eeq
comes from the contribution of $\tdeltag$, and
\beq
F_v(k) = f(\Omega_{\rm M})\frac{1}{\bar{\rho} } \int dM \frac{dn}{dM} M b(M) R_1(M) {\tu_m(k|M)}
\eeq
comes from the contribution of $\tdelta$.  Here $\tu_m(k|M)$ denotes
the {\it dark matter} density profile normalized the same way as
$\tug$, and $\bar{\rho}$ is the average matter density of the
universe.  We assume that $\tu_m(k|M)$ follows the
Navarro--Frenk--White (NFW) profile \citep{Navarro97} throughout the
paper.

The left-hand panel of Fig.~\ref{fig:PS_rsd} shows an example of the
contribution to the total galaxy power spectrum by the one-halo (blue)
and two-halo (red) terms.  The input of halo model will be detailed in
Section \ref{sec:assumption}.  The solid and dashed curves correspond
to including and excluding the effect of RSD.  As can be seen,
including RSD significantly reduces the power at small scale.  We also
note that the scale where one- and two-halo terms cross shifts very
slightly due to RSD.

The right-hand panel of Fig.~\ref{fig:PS_rsd} presents the comparison
between our model and one of the power spectra from the WiggleZ
survey, provided by \citet{Parkinson12}.  The green dashed/blue solid
curve corresponds to the theoretical $P(k)$ before/after convolving
with the window function of WiggleZ.  We assume that the HOD is
described by the five parameters in equation~(\ref{eq:HOD}); we fit
for these five parameters and show the model corresponding to the
best-fitting parameters.  This figure is only for the purposes of
illustration; details of the fitting procedure will be presented in a
future paper.

\section{Baseline model and fiducial dark energy constraints}\label{sec:fiducial}

In this section, we describe our inputs for the halo model,
assumptions about the survey, predictions for the galaxy power
spectrum, and Fisher matrix calculations of the statistical and
systematic errors.

\subsection{Baseline assumptions}\label{sec:assumption}
We use the virial mass $\Mvir$ of dark matter haloes throughout this work
and adopt the following functions in our halo model calculations: 
\begin{itemize}

\item Mass function (${\rm d}n/{\rm d}M$) and halo bias ($b(M,z)$):
based on the fitting functions in \citet{Tinker08,Tinker10}, which are
derived from $N$-body simulations and can achieve approximately 5 per
cent accuracy for the mass function and 6 per cent for the halo bias.

\item Density profile: based on the universal NFW profile \citep{Navarro97}, which is described by one
concentration parameter $c_{\rm vir}$
\beqa u_{\rm NFW}(r|\Mvir) &\propto& \frac{1}{(r/r_s)(1+r/r_s)^2} \ , \\ 
c_{\rm vir} (\Mvir)&=& \Rvir / r_s
  \ . \nonumber 
\eeqa

\item Concentration--mass relation: based on the relation in
\citet{Bhattacharya11}, which will be further discussed in Section
\ref{sec:c-M}.  In the presence of significant scatter in the $c$--$M$
relation, we perform the integration
  \beq
  \begin{aligned}
  u(r|\Mvir) = \int d\cvir P(\cvir|\Mvir)u(r|\Mvir(\cvir)) \ .
  \end{aligned}
  \eeq
Throughout this paper, we assume that $\cvir$ has a Gaussian
distribution for a given $\Mvir$ with a scatter of 0.33, based on the
finding of \cite{Bhattacharya11}.

\item Velocity dispersion: based on the scaling relation between dark
matter velocity dispersion and halo mass from
\citet{Evrard08}
\beq
\sigmav^{\rm DM}= 1082.9\left(\frac{h(z)M_{200}}{10^{15} M_\odot}\right)^{0.3361}  \ {\rm km\ s^{-1}}\ . 
\eeq
We convert the mass $M_{200}$ to $\Mvir$ based on
\cite{HuKravtsov03}. Since the scatter in the velocity dispersion is
expected to be small (4 per cent), it is not included in our
calculation.

\item HOD: based on the parametrization from
  \citet{Zheng05} and the fiducial parameters from \citet{Coupon12}, both
  of which will be discussed in detail in Section \ref{sec:HOD}.

\end{itemize}

We assume a fiducial galaxy survey covering $\fsky=1/3$ of the full
sky (about 14 000 square degrees), similar to the BigBOSS
experiment\footnote{\tt http://bigboss.lbl.gov/}.  We assume that the
survey depth is comparable to the Canada--France--Hawaii Telescope
Legacy Survey (CFHTLS) results presented in \cite{Coupon12};
specifically, we assume five redshift bins in the range $0.2<z<1.2$,
and the limiting magnitude in each bin is summarized in
Table~\ref{tab:HOD}.  We assume no uncertainties in the redshift
measurements of galaxies.  Given that the assumption of such a deep,
wide spectroscopic survey {\it may} be somewhat optimistic, our
required control of systematic errors may be somewhat more stringent
than what BigBOSS needs.

We include seven cosmological parameters, whose fiducial values are
based on the {\em Wilkinson Microwave Anisotropy Probe} 7 constraints
\citep{Komatsu11}: total matter density relative to critical
$\Omega_{\rm M} = 0.275$; dark energy equation of state today and its
variation with scale factor $w_0 = -1$ and $w_a = 0$ respectively;
physical baryon and matter densities $\Omega_{\rm b} h^2 = 0.02255$
and $\Omega_{\rm M} h^2 = 0.1352$; spectral index $n_s=0.968$; and the
amplitude of primordial fluctuations $A = \Delta_\zeta^2
(k=0.002\hiMpc) = 2.43\times10^{-9}$. We assume a flat universe; thus,
dark energy density $\Omega_{\rm DE} = 1-\Omega_{\rm M}$.

\subsection{Likelihood function of $P(k)$ and error forecasting}

Here we follow the derivations in \cite{Scoccimarro99} and
\cite{CoorayHu01} but use a different convention for the Fourier
transform (see Appendix~\ref{app:halomodel}).  If we assume a thin
shell in $\ln k$ space with width $\delta \ln k$ around $\ln k_i$, the
power spectrum estimator reads
\beq
\hat P(k_i) =\int_{k_i} \frac{{\rm d}^3 \vk}{V_s(k_i)}\delta(\vk)\delta(-\vk) +\frac{1}{\brhog}\ ,
\eeq
where
\beq
V_s (k_i) = 4{\rm\pi} k_i^3 \delta\ln k  \ ,
\eeq
and $1/\brhog$ accounts for the effect of shot noise. The first term
of $\hat{P}$ is calculated based on the halo model results described
in Section~\ref{sec:Pk_rsd}.

The covariance of power spectrum is given by
\beq\bal
C_{ij} &\equiv  \avg{\hat P(k_i) \hat P(k_j)} - \avg{\hat P(k_i)} \avg{\hat P(k_j)} \\[0.2cm]
&\quad =\frac{(2{\rm\pi})^3}{V_z}\frac{  2 P(k_i)^2}{V_s(k_i)} \delta_{ij}  + \bar T_{ij}
  \ , \label{eq:C_ij}
\eal\eeq
where the second term on the right-hand side is the contribution from
the connected term given by the trispectrum describing the
non-Gaussian nature of the random field
\beq \bar T_{ij} =
\int_{k_i} \frac{{\rm d}^3 \vk_1}{V_s(k_i)}\int_{k_j} \frac{{\rm d}^3 \vk_2}{V_s(k_j)}
T(\vk_1,-\vk_1,\vk_2,-\vk_2) \ .
\label{eq:Tij}
\eeq
We provide the detailed derivation in Appendix~\ref{app:Cij}.  In
equation~(\ref{eq:C_ij}) $V_z$ is the volume of the redshift bin, $V_z
= \Omega_{\rm survey}\int r^2(z)/H(z) dz$, where the integral is
performed over the redshift extent of the bin.

The calculation of $\bar T_{ij}$ involves four-point statistics, which
is non-trivial to calculate.  Fortunately, \citet{CoorayHu01} have
shown that only the one-halo term dominates at the scale where the
contribution of $\bar T_{ij}$ to $C_{ij}$ is not negligible;
therefore, we only need to calculate the one-halo contribution:
\beq\bal
T^{\rm 1h}(k_1, k_2, k_3, k_4) &= \frac{1}{V_z \brhog^4}\int dM \frac{dn}{dM} \\
&\times\avg{\displaystyle{\left . N\choose4\right |}M} f(k_1, k_2, k_3, k_4;M)
\label{eq:T1h}
\eal\eeq
where 
\begin{equation}
\begin{aligned}
&\avg{\displaystyle{\left . N\choose4\right |}M} f(k_1, k_2, k_3, k_4;M) \\
&=\avg{\displaystyle{\left . \Ns\choose3\right |} M } \frac{1}{4}\left( {\rm\pi}_{i=1}^{3} \tu(k_i|M) + \mbox{cyc.}\right) \\
&+\avg{\displaystyle{\left . \Ns\choose4\right |} M } {\rm\pi}_{i=1}^{4} \tu(k_i|M) \ .
\end{aligned}
\end{equation}
Analogous to the case of $P_{\rm gg}^{\rm 1h}$ considered in
Section~\ref{sec:Pk_basic}, the first term accounts for quadruplets
composed of one central and three satellite galaxies, and the second
term accounts for the quadruplets composed of four satellite galaxies.
We assume that $P(\Ns|M)$ follows the Poisson distribution so that
$\avg{\textstyle{\Ns\choose3}|M}=\avg{\Ns|M}^3/3!$ and
$\avg{\textstyle{\Ns\choose4}|M}=\avg{\Ns|M}^4/4!$.

We employ the Fisher matrix formalism to forecast the statistical
errors of the cosmological and nuisance parameters based on the
fiducial survey.  The Fisher matrix reads
\beq
\begin{aligned}
&F_{\alpha\beta}  
= \sum_z
\sum_{i,j} \frac{\partial P_i}{\partial \theta_\alpha}
\left[ \frac{(2{\rm\pi})^3}{V_z} \frac{P_i^2}{2{\rm\pi} k_i^3 \delta\ln k} \delta_{ij}+ {T_{ij}} \right]^{-1}
\!\!\frac{\partial P_j}{\partial \theta_\beta} \nonumber\\[0.10cm]
&= \sum_z
\sum_{i,j} \frac{\partial \ln P_i}{\partial \theta_\alpha}
\left[ \frac{(2{\rm\pi})^3}{V_z} \frac{1}{2{\rm\pi} k_i^3 \delta\ln k} \delta_{ij}+ \frac{T_{ij}}{P_i P_j} \right]^{-1}
\!\!\frac{\partial \ln P_j}{\partial \theta_\beta} \ , \nonumber
\end{aligned}
\eeq
where $\alpha$ and $\beta$ are indices of model parameters, while $i$
and $j$ refer to bins in wavenumber which have a constant logarithmic
width $\delta \ln k$ and extend out to the maximum wavenumber $\kmax$.
We adopt $\delta\ln k$ = 0.1, which has been tested to be small enough
to ensure convergence.  The best achievable error in the parameter
$\theta_\alpha$ is given by
\beq
\sigma_{\theta_\alpha} = [({\mathsfbf F}^{-1})_{\alpha\alpha}]^{1/2} .
\eeq

Throughout this work, unless otherwise indicated, the full set of
parameters considered is given by
\beq\bal
\theta_{\rm full} &= (w_0, w_a, \Omega_{\rm DE}, \Omega_{\rm M} h^2, \Omega_{\rm b} h^2, n_s, \ln A; \\
&\log_{10}M_{\rm min},  \sigma_{\log_{10} M}, \log_{10} M_0, \log_{10} M_1,  \alpha_{\rm sat}) \ .
\eal\eeq
The first seven are the cosmological parameters introduced in
Section~\ref{sec:assumption}, while the last five are the nuisance
parameters describing the HOD and will be discussed in
Section~\ref{sec:Zheng}.

\begin{figure}
\includegraphics[width=\columnwidth]{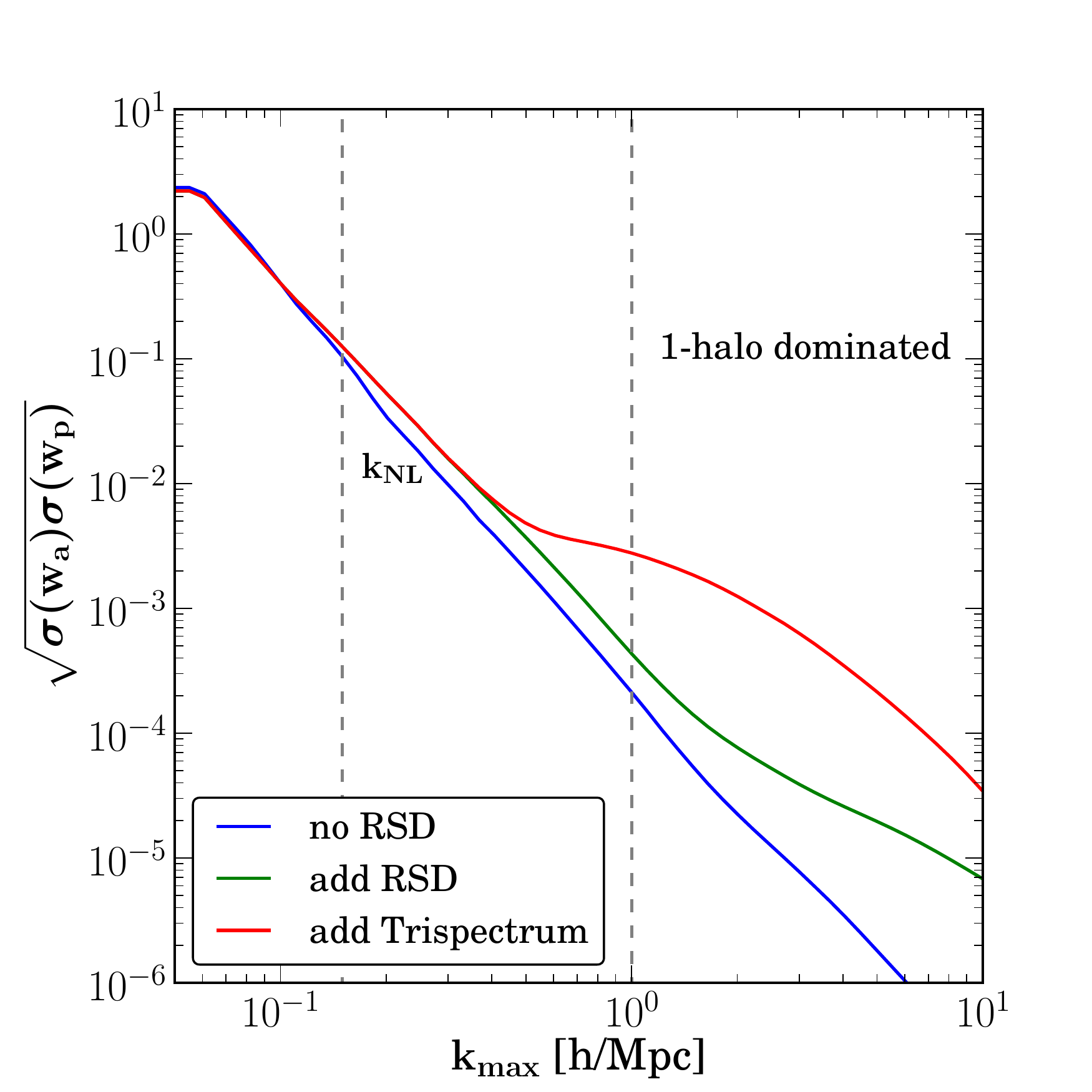}
\caption{Dark energy information content from $P(k)$, based on our
fiducial survey assumptions.  The $x$-axis corresponds to the largest
$k$ (smallest scale) assumed to be reliably measured and interpreted.
The blue curve corresponds to a Gaussian likelihood function and
assumes no RSD; it leads to unrealistically tight constraints for
large $\kmax$.  The green curve includes RSD, and the red curve
further includes the trispectrum correction to the covariance matrix.
As can be seen, including these two effects reduces the small-scale
information.}
\label{fig:sigmaw_with_Cij}
\end{figure}

\subsection{Fiducial constraints without systematics}

To represent the statistical power of an upcoming galaxy redshift
survey, in the limiting case of no nuisance parameters, we consider
the inverse of the square root of the dark energy figure of merit,
originally defined as the inverse of the forecasted 95 per cent area
of the ellipse in the $w_0$--$w_a$ plane \citep{HutererTurner01,DETF}.
In other words, our parameter of interest is
$\sqrt{\sigma(w_a)\sigma(w_p)}$, where $w_p$ is the pivot that
physically corresponds to $w(a)$ evaluated at the scale factor where
the constraint is the best. This quantity takes into account the
temporal variation of dark energy, and the square root serves to
compare it fairly to the constant $w$; the two quantities, $\sigma(w)$
and $\sqrt{\sigma(w_0)\sigma(w_p)}$, tend to show very similar
behaviour.  For our fiducial survey, the statistical error in our
parameter combination of interest is $\sqrt{\sigma(w_a)\sigma(w_p)}=$
0.4 (or 0.003) for $\kmax = 0.1$ (or 1) $\hoverMpc$, without external
priors. When we add the Planck Fisher matrix (Hu, private
communication), $\sqrt{\sigma(w_a)\sigma(w_p)}$ becomes 0.002 (or
0.0002) for $\kmax = 0.1$ (or 1) $\hoverMpc$.

Fig.~\ref{fig:sigmaw_with_Cij} presents the expected dark energy
constraints as a function of $\kmax$, without nuisance parameters or
systematic errors for the moment, for three levels of sophistication
in the theory.  We proceed in steps: the blue curve corresponds to no
RSD (Section \ref{sec:Pk_basic}) with a Gaussian likelihood function.
In this case, the dark energy constraints increase sharply with
$\kmax$, indicating that these assumptions are unrealistic.  The green
curve includes the RSD (Section \ref{sec:Pk_rsd}), which reduce the
dark energy information from small scales.  The red curve further
includes the effect of non-Gaussian likelihood [$\bar{T}_{ij}$ from
equation~(\ref{eq:Tij})], which reduces the information at high $k$
even more.

\subsection{Systematic bias in model parameters}

In this work, we estimate the systematic shifts in parameter inference
caused by using an inadequate model.  In particular, if we assume a
problematic model that produces a power spectrum $P_{\rm sys}(k)$ that
systematically deviates from the truth $P_{\rm fid}(k)$, we will
obtain parameters that systematically deviate from their true values:
$\theta_{\rm sys} = \theta_{\rm fid} + \Delta \theta$.  The systematic
shifts in parameters can be obtained through a modified Fisher matrix
formalism \citep{Knox98}:
\beq
\Delta \theta_\alpha = \sum_\beta ({\mathsfbf F}^{-1})_{\alpha\beta} G_{\beta} \, ,
\eeq
where
\begin{equation}
\begin{aligned}
G_{\beta} & \equiv \sum_z
\sum_{i,j} (\ln P_{\rm sys, \it i}- \ln P_{\rm fid, \it i})\\
&\times\left[ \frac{(2{\rm\pi})^3}{V_z}\frac{1}{2{\rm\pi} k_i^3 \delta\ln k} \delta_{ij}+ \frac{T_{ij}}{P_iP_j} \right]^{-1}
\frac{\partial \ln P_j}{\partial \theta_\beta} \, .
\end{aligned}
\end{equation}
To determine the significance of systematic errors, we calculate the
systematic shifts $\Delta\chi_{\rm tot}^2$ in the full
high-dimensional parameter space,
\beq
\Delta\chi^2_{\rm tot} = \Delta \theta^T {\mathsfbf F} \Delta \theta \, ,
\label{eq:Delta_chi2}
\eeq 
where $\Delta\theta$ is the vector of the systematic shifts of
parameters.  Both $\Delta\theta$ and the Fisher matrix ${\mathsfbf F}$
include cosmological and nuisance parameters. The systematic bias is
considered significant if the inferred $\theta_{\rm sys}$ lies outside
the 68.3 per cent confidence interval of the Gaussian likelihood
function centred on $\theta_{\rm fid}$; in other words, the bias is
`greater than the 1 $\sigma$ dispersion'.  For example, in a full
12-dimensional parameter space considered here, the 68.3 per cent
confidence interval corresponds to $\Delta\chi_{\rm tot}^2 = 13.7$.

\section{Self-calibration of HOD parameters}\label{sec:HOD}
\begin{figure}
\includegraphics[width=\columnwidth]{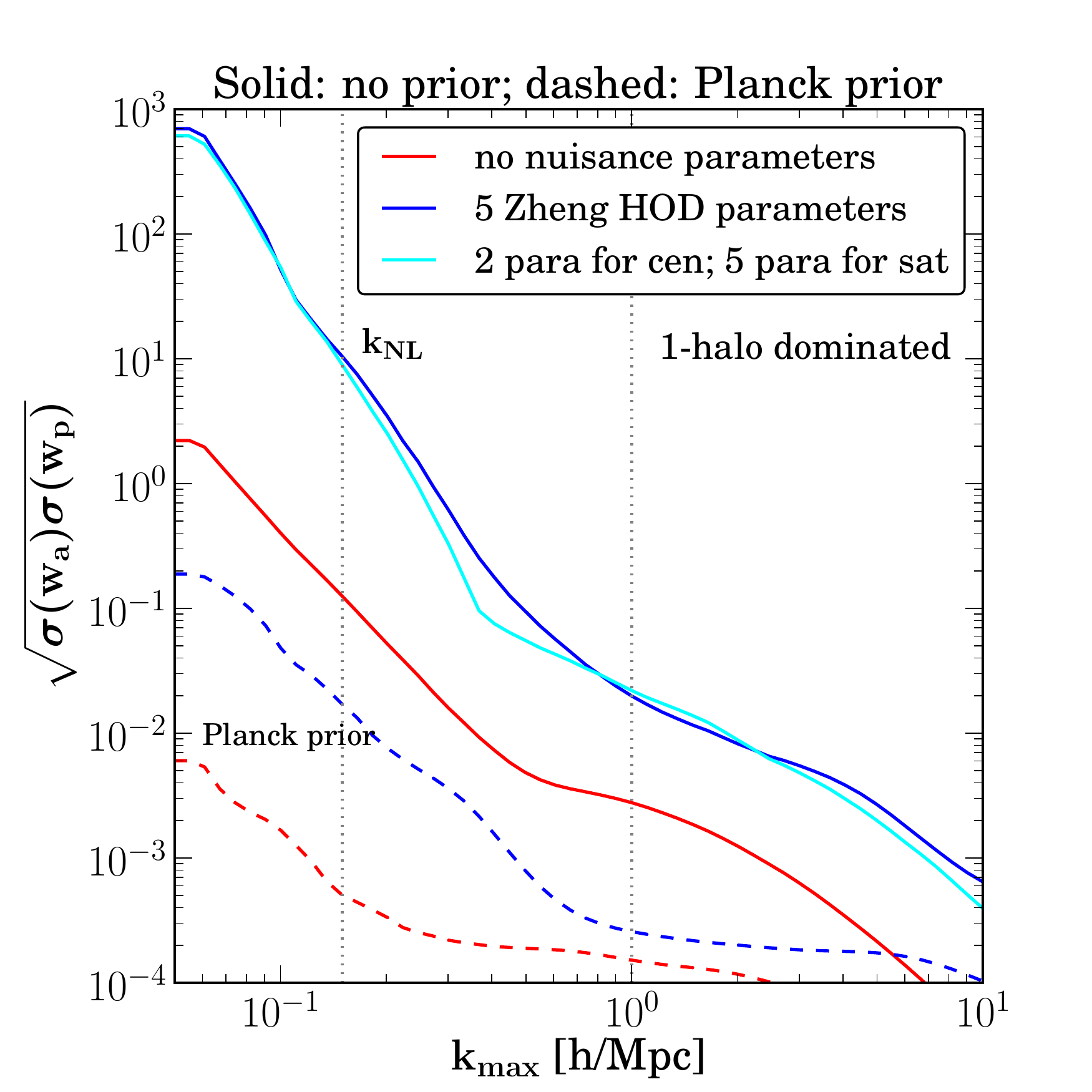}
\caption{ Self-calibration of HOD parameters.  We show the dark energy
constraints as a function of the highest $k$ used in the survey. The
red curve corresponds to no nuisance parameters.  The dark blue curve
corresponds to five nuisance parameters based on the parametrization
in \protect\cite{Zheng05}, while the cyan curve corresponds to a
piecewise continuous parametrization for satellite galaxies, with one
parameter in each of the five mass bins.  Including nuisance
parameters in either parametrization systematically increases the dark
energy uncertainties by one or two orders of magnitude. The dashed
curves include the Planck prior and assume the same nuisance
parameters as their solid-curve counterparts.  }
\label{fig:HOD_nuisance}
\end{figure}

In this section, we focus on the efficacy of self-calibrating the HOD
parameters, that is, determining these parameters from the survey
concurrently with cosmological parameters.  Since these HOD parameters
are not known a priori, one usually marginalizes over them along with
cosmological parameters \cite[e.g.,][]{Tinker12}, which inevitably
increases the uncertainties in cosmological parameters.  Here we focus
on the statistical uncertainties and assume no systematic error; in
the next section, we will compare these statistical errors with
systematic shifts of parameters.

We focus on two parametrizations of HOD: one is based on
\citet{Zheng05} and the other is based on a piecewise continuous
parametrization.

\subsection{Zheng et al.~ parametrization}\label{sec:Zheng}
\begin{table*}
\centering
\setlength{\tabcolsep}{0.5em}
\begin{tabular}{|c|c||c|c|c|c|c|}
\hline
\rule[-2mm]{0mm}{6mm}
Redshift & $M_g-5\log_{10}h$ & $\log_{10}{M_{\rm min}}$ & $\log_{10}{M_1}$&$\log_{10}{M_{0}}$ &  $\sigma_{\log_{10} M}$ & ${\alpha_{\rm sat}}$ \\\hline 
\rule[-2mm]{0mm}{6mm}$0.2 < z < 0.4$ & -17.8 & 11.18 & 12.53 &   7.54 & 0.40 & 1.10 \\ \hline
\rule[-2mm]{0mm}{6mm}$0.4 < z < 0.6$ & -18.8 & 11.48 & 12.66 & 10.96 & 0.43 & 1.09 \\ \hline
\rule[-2mm]{0mm}{6mm}$0.6 < z < 0.8$ & -19.8 & 11.77 & 12.83 & 11.54 & 0.50 & 1.07 \\ \hline
\rule[-2mm]{0mm}{6mm}$0.8 < z < 1.0$ & -20.8 & 12.14 & 13.21 & 12.23 & 0.35 & 1.12 \\ \hline
\rule[-2mm]{0mm}{6mm}$1.0 < z < 1.2$ & -21.8 & 12.62 & 13.79 &   8.67 & 0.30 & 1.50 \\ \hline
\end{tabular}
\caption{Fiducial values for the HOD parameters, adopted from
  \citet{Coupon12} based on CFHTLS.}
\label{tab:HOD}
\end{table*}

The HOD describes the probability distribution of having $N$ galaxies
in a halo of mass $M$.  In principle, the HOD is specified by the full
distribution $P(N|M)$; in practice, modelling of the two-point
statistics only requires $\avg{\Nc|M}$, $\avg{\Ns|M}$, and
$\avg{\Ns(\Ns-1)|M}$.  We follow the HOD parametrization from
\citet{Zheng05}, which separates the contribution from central and
satellite galaxies:
\beqa
\avg{\Nc|M }&=& \frac{1}{2}\left[1+{\erf}\left( \frac{\log_{10} M - \log_{10} M_{\rm min}}{\sigma_{\log_{10} M}} \right)\right] \\
\avg{\Ns|M }&=& \avg{\Nc|M }\times\left( \frac{M-M_0}{M_1} \right)^{\alpha_{\rm sat}} \label{eq:HOD}
\eeqa
The first equation describes the contribution from the central galaxy;
$M_{\rm min}$ corresponds to the threshold mass where a halo can start
to host a galaxy that is observable to the survey, and
$\sigma_{\log_{10}M}$ describes the transition width of this
threshold.  The second equation describes the contribution from
satellite galaxies, whose number is assumed to follow a power law, and
$M_0$ is the cutoff mass.  In addition, we make the widely-adopted
assumption that $P(\Ns|M)$ follows a Poisson distribution, i.e.,
\beq 
\avg{\Ns(\Ns-1)|M} = \avg{\Ns|M }^2 \ .  
\eeq

We adopt the fiducial values from \citet{Coupon12}, which are
constrained using the projected angular two-point correlation function
$w(\theta)$ from the CFHTLS out to $z$ = 1.2.  We use the same binning
and limiting magnitude as in \cite{Coupon12}; the values are
summarized in Table~\ref{tab:HOD}.  We do not use the error bars
quoted there as our priors because we would like all parameters to be
self-calibrated consistently.

Under these assumptions, we have five nuisance parameters
$(\log_{10}M_{\rm min},\ \sigma_{\log_{10} M},\ \log_{10} M_0,\
\log_{10} M_1,\ \alpha_{\rm sat})$ for each of the five redshift bins,
i.e., 25 parameters in total. We assume that each of the five distinct
nuisance parameters varies coherently across the five redshift bins,
and is therefore described by a single parameter.  Under this
assumption, instead of 25 nuisance parameters, we only use five
nuisance parameters to describe the uncertainties of all HOD
parameters.  We parametrize the variations around the fiducial values:
\beq
\theta^{\rm HOD}_{i} (z) = h_i \theta^{\rm HOD,fid}_i (z) \quad (i = 1,...,5) \ ,
\eeq 
where $h_i$ are the dimensionless parameters describing the
uncertainties of the aforementioned 5 HOD parameters.  We note that
this choice of five HOD parameters only represents one possible model;
depending on the data available and the astrophysical motivation, in
principle one can use a more general model to describe the evolution
of HOD.  Increasing the number of degrees of freedom describing the
evolution of HOD will inevitably lead to degradation in the dark
energy constraints, and it will be very important to establish the
total number of degrees of freedom necessary to model the HOD and its
uncertainties.

We explore how well these parameters can be self-calibrated by $P(k)$
without the aid of priors.  Fig.~\ref{fig:HOD_nuisance} shows the dark
energy constraints as a function of $\kmax$, with fixed nuisance
parameters (red) and with these 5 marginalized nuisance parameters
(dark blue).  The RSD and the full covariances of $P(k)$ are included
in this calculation.  Clearly, the dark energy constraints are
weakened by approximately about one or two orders of magnitude when we
marginalize over HOD parameters.

\subsection{Piecewise continuous parametrization of HOD parameters}\label{sec:piecewise}
\begin{figure*}
\includegraphics[width=2.2\columnwidth]{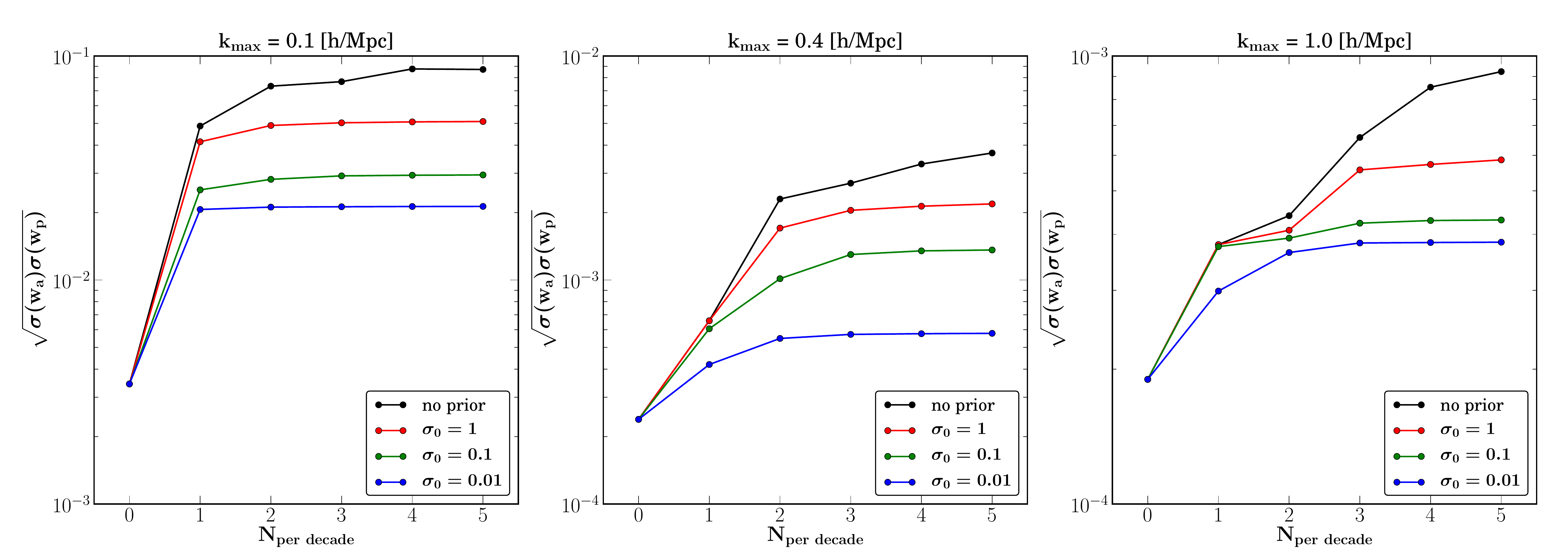}
\caption{Dark energy constraints with self-calibrated piecewise
continuous HODs. The three panels correspond to $\kmax$ = 0.1, 0.4 and
1 $\hoverMpc$.  The $x$-axis corresponds to the number of parameters
used to describe $\avg{\Ns|M}$ per decade of mass, and the $y$-axis
corresponds to the dark energy constraints.  The black curve
corresponds to no prior, and the constraints are degraded with larger
number of parameters.  The other curves correspond to consistently
adding {\it a fixed total} prior per decade of mass; that is,
$\sigma_{f_i} = \sigma_0 \sqrt{N_{\rm per\ decade}}$\ , where
$\sigma_0$ = 1, 0.1, or 0.01.  We note that one parameter per decade
is sufficient for $\kmax$ = 0.1 $\hoverMpc$, while two or three
parameters are needed for higher $\kmax$.  Note that for higher
$\kmax$, the HOD parameters are better self-calibrated, and the dark
energy constraints are less dependent on the prior on HOD nuisance
parameters.  }
\label{fig:piecewise_Npara}
\end{figure*}

One potential worry with the parametrization in
equation~(\ref{eq:HOD}) is whether $\avg{\Ns|M}$ is accurately
described by a power law.  To address this, we propose a less
model-dependent, piecewise continuous parametrization for
$\avg{\Ns|M}$. We divide the halo mass range into $n_{\rm bins}$ bins
and assign a parameter describing the uncertainties of HOD in each
bin.  That is,
\beq
\avg{\Ns|M} = \sum_{i=1}^{n_{\rm bins}}   \Theta_i(M) f_i \avg{\Ns|M}_{\rm fid}  \ ,
\eeq
where $\Theta_i(M)$ defines the binning and equals 1 in
$[M_i, M_{i+1}]$ and 0 elsewhere, while $f_i$ is the free parameter in
bin $i$ and describes the uncertainty of $\avg{\Ns|M}$ in this bin.

We still assume $P(\Ns|M)$ to be a Poisson distribution, which now implies
\begin{equation}
\begin{aligned}
\avg{\Ns(\Ns-1) |M} &= \avg{\Ns|M} ^2 \\[0.2cm]
 &= \sum_{i=1}^{n_{\rm bins}} \Theta_i(M) f_i^2 \avg{\Ns|M}^2_{\rm fid}.
\end{aligned}
\end{equation}
We start with one parameter per decade in mass, using $n_{\rm bins}=5$
parameters between $10^{11}$ and $10^{16} \, \hiMsun$, equally spaced
in $\log_{10}M$.  We assume these parameters to be independent of
redshift.  The cyan curve in Fig.~\ref{fig:HOD_nuisance} corresponds
to marginalizing over these five piecewise continuous parameters for
$\avg{\Ns|M}$ and two parameters ($\log_{10}M_{\rm min}$,
$\sigma_{\log_{10}M}$) for $\avg{\Nc|M}$, with no prior on them.

Fig.~\ref{fig:piecewise_Npara} shows the dependence of dark energy
constraints on the number of parameters describing $\avg{\Ns|M}$ per
decade of mass, $N_{\rm per\ decade}$.  The three panels correspond to
$\kmax$ = 0.1, 0.4, and 1 $\hoverMpc$.  The Planck prior is included
in this calculation.  The black curve corresponds to no prior on $f_i$
and shows strong degradation with increasing $N_{\rm per\ decade}$ as
one would expect.  When $\kmax$ is small, the prior knowledge of HOD
is important to improve the dark energy constraints.  On the other
hand, when $\kmax$ is large, HOD can be well self-calibrated, and the
prior is not as important.

To enable a fair comparison of priors, however, we would like to increase
the freedom in the HOD model while fixing the overall uncertainty per
decade. To do this, we impose a fixed prior {\it per decade} of mass:
\beq
\sigma_{f_i} = \sigma_{0}\sqrt{N_{\rm per\ decade}} \label{eq:sigma_fi}
\eeq
so that the total prior per unit $\log_{10}M$, when we add the Fisher
information from all $f_i$, is $\sigma_0$ regardless of the value of
$N_{\rm per\ decade}$.

The red/green/blue curves in Fig.~\ref{fig:piecewise_Npara} correspond
to imposing $\sigma_0=1/0.1/0.01$.  For $\kmax = 0.1\, \hoverMpc$, the
dark energy constraints converge when we use one parameter per decade
of mass regardless of the prior on nuisance parameters.  When $\kmax >
0.1\, \hoverMpc$, a few more parameters per decade in mass are
required for the results to converge. For example, for $\kmax$ = 0.4
(1.0) $\hoverMpc$, we need two (three) parameters per decade to ensure
convergence.  The required number of parameters also somewhat depends
on the prior.

We note that the HOD parameters are progressively better
self-calibrated when we go to higher $\kmax$; when $\kmax = 1\,
\hoverMpc$, self-calibrating the five HOD parameters only moderately
degrades the dark energy constraints.  This finding encourages future
surveys to further push towards high $\kmax$ for rich cosmological and
astrophysical information.

\section{Systematic errors due to the uncertainties in halo mass function}\label{sec:MF}

\begin{figure}
\includegraphics[width=\columnwidth]{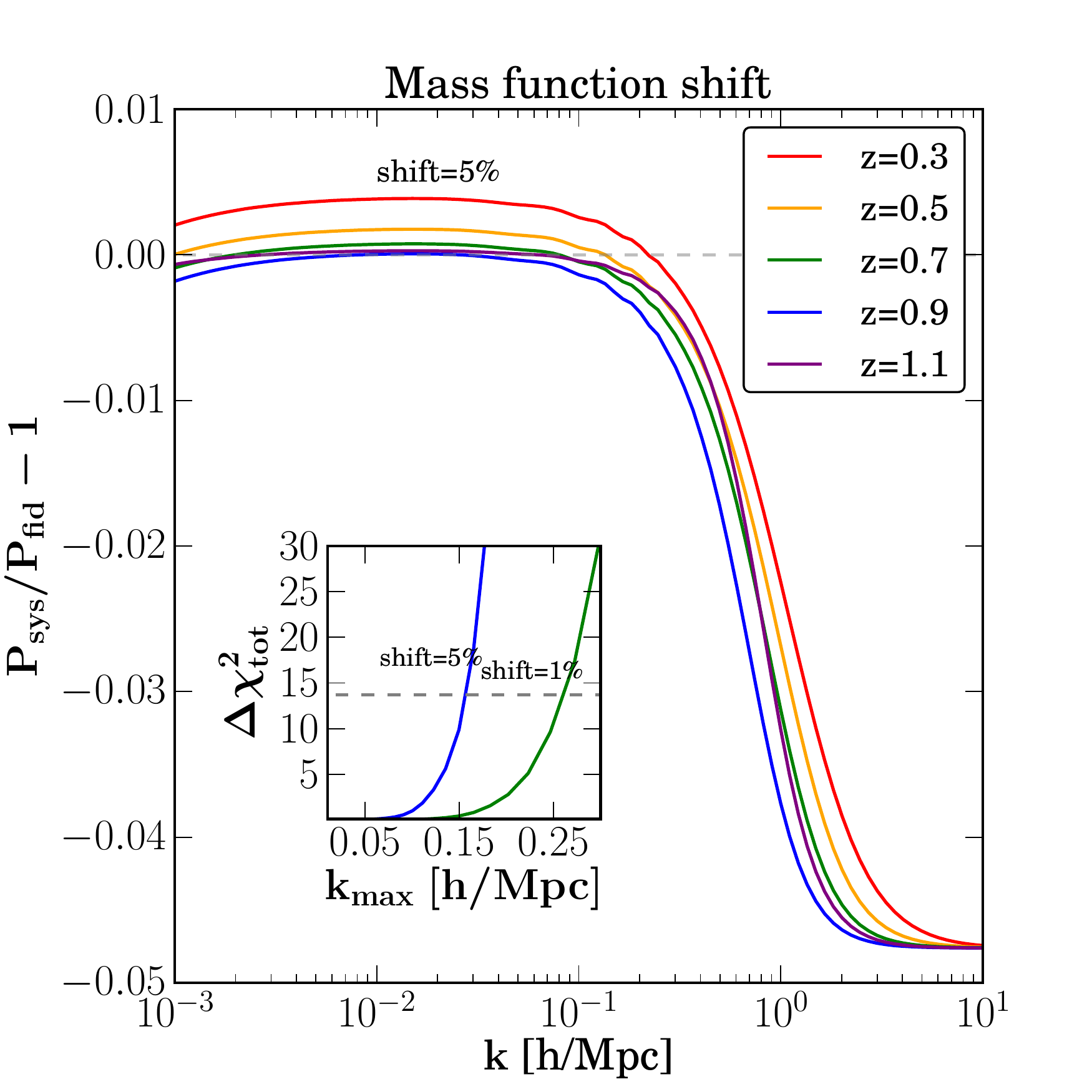}
\caption{ Impact of the uncertainty in the halo mass function on
$P(k)$.  The main panel shows the systematic shifts of $P(k)$ when the
mass function is shifted by a constant 5 per cent (independent of mass
and redshift).  The inset shows $\Delta\chi^2_{\rm tot}$ as a function
of $\kmax$ when the mass function is shifted by 1 or 5 per cent.  The
horizontal dashed line marks $\Delta\chi^2_{\rm tot} = 13.7$, the
1-$\sigma$ deviation in the 12-dimensional parameter space.  As can be
seen, 5 per cent (1 per cent) allows $\kmax$ up to 0.15 (0.25)
$\hoverMpc$.}
\label{fig:MF}
\end{figure}

In this section, we focus on the effect of the uncertainties in the
halo mass function on the cosmological constraints from galaxy
clustering.  The mass function has been widely explored analytically
\cite[e.g.,][]{PressSchechter74} as well as numerically using dark
matter $N$-body simulations \citep[e.g.,][]{ShethTormen99,Sheth01,
Jenkins01,Evrard02,Reed03,Warren06,Lukic07,
CohnWhite08,Tinker08,Lukic09,Crocce10,Bhattacharya11,Reed13,Watson13}
and hydrodynamical simulations \citep[e.g.,][]{Rudd08,Stanek09,Cui12}.
The different fitting formulae for the mass function are often based
on different halo identification methods and mass definitions;
therefore, instead of drawing a direct comparison between different
fitting formulae, we choose one specific fiducial model and explore
the uncertainties relative to this model.\footnote{It has been shown
that for surveys of cluster abundance such as the Dark Energy Survey,
$\sim1$ per cent accuracy in mass function is required to avoid
significant degradation in dark energy constraints
\cite[see][]{Cunha09MF,Wu09b}.  Here we would like to explore whether
the same accuracy is sufficient for surveys of galaxy clustering. }

We use the fitting function from \citet[][described in
Section~\ref{sec:assumption}]{Tinker08}, which has been calibrated
based on a large suite of simulations implementing different $N$-body
algorithms and different versions of $\Lambda$CDM cosmology;
therefore, it is likely to fairly represent the uncertainties in the
mass function calibration.  \cite{Tinker08} quoted a statistical
uncertainty of $\lesssim5$ per cent at $z=0$ ($\sim1$ per cent around
$M_*$).  However, the uncertainties are presumably larger at higher
redshift and can further increase if the effects of baryons are taken
into account.

We explore the effect of a small constant shift of the halo mass
function, parametrized as
\beq
\left(\frac{dn}{dM}\right)= (1+\epsilon)
\left(\frac{dn}{dM}\right)_{\rm fid} \ .
\eeq
The main panel of Fig.~\ref{fig:MF} shows the impact of $\epsilon =
0.05$ on $P(k)$.  For the two-halo term (small $k$), $P(k)$ changes by
less than 1 per cent, because a constant shift in the mass function
only affects $F_v$ describing the large-scale RSD (see
equation~\ref{eq:P2h_rsd}).  For the one-halo term (large $k$), $P(k)$
changes by $-5$ per cent, which can be easily seen from
equation~\ref{eq:P1h_rsd}; the numerator includes one integration of
${\rm d}n/{\rm d}M$ (galaxy pairs in one halo) while the denominator
includes the square of such an integration.

We next see how this systematic shift in $P(k)$ impacts cosmological
parameters.  We use $\Delta\chi^2_{\rm tot} = 13.7$ [1 $\sigma$ errors
in a 12-dimensional parameter space; see equation (\ref{eq:Delta_chi2})]
as our criterion of significant impact from the systematic error. We
calculate $\Delta\chi^2_{\rm tot}$ using the Fisher matrix for seven
cosmological parameters (Section~\ref{sec:assumption}) and five HOD
parameters (Section~\ref{sec:Zheng}). Throughout this and the next
section, we use the Planck prior but {\it no} priors on HOD
parameters.  We believe that these two assumptions reflect reality in
the next 5--10 years, when Planck data will firmly pin down certain
combinations of cosmological parameters, while the determination of
the nuisance HOD quantities will still be in flux.  We note that
unbiased priors always {\it decrease} the resulting systematic bias
(for a proof, see appendix A of \citealt{BernsteinHuterer10}) and make
the theoretical requirements less stringent.  Thus, any prior on HOD
parameters will alleviate the systematic biases and make the required
accuracy of theory less stringent.

The inset of Fig.~\ref{fig:MF} shows how $\Delta\chi^2$ depends on
$\kmax$, for 5 per cent (blue) and 1 per cent (green) systematic
shifts in the mass function.  As can be seen, a 5 per cent (1 per cent) 
shift in the mass function can cause a significant systematic error
at $\kmax=0.15$ (0.25) $\hoverMpc$.  We note that at these scales,
$P(k)$ is still dominated by the two-halo term; therefore, the
systematic shifts caused by the mass function are mainly related to
the large-scale redshift distortion (the Kaiser effect).  However,
this large-scale effect can be mitigated by prior knowledge of
$\sigma_8$ (e.g., $\sigma_8$ constraints from galaxy cluster counts;
\citealt{Rozo10}), with which the large-scale galaxy bias can be
calibrated.  Therefore, by calibrating the large-scale clustering
amplitude, one can in principle reduce the impact of the uncertainties
in the mass function.

Finally, we note that the halo bias $b(M)$ is also currently being
actively studied \citep[e.g.,][]{Tinker10,Ma11,Manera11,Paranjape13}.
The uncertainty in the halo bias is related to the uncertainty in the
mass function; for example, \cite{Tinker10} have indicated that their
fitting function for the halo bias has an $\sim6$ per cent
uncertainty, which is related to the uncertainty of their mass
function.  In addition, the uncertainties and systematics in $b(M)$
will lead to a constant shift in the two-halo term (see
equation~\ref{eq:P2h_rsd}).  In this case, holding the galaxy bias
fixed will cause a huge systematic shift in cosmological parameters
(for example, $\sigma_8$); therefore, it is necessary to fit the
overall galaxy bias to the large-scale clustering data.  In this work,
we do not specifically explore the impact of uncertainties of the halo
bias because the halo bias determines the large-scale clustering
amplitude, which can be observationally calibrated when combined with
independent knowledge of $\sigma_8$.  On the other hand, we note that
a scale-dependent bias can arise from the primordial non-Gaussianity
\citep[e.g.,][]{Dalal08} or small-scale non-linearity
\citep[e.g.,][]{Smith07}.  In this case, one could resort to multiple
tracers of large-scale structure \citep[e.g.,][]{Seljak09,Cacciato12},
knowledge of primordial non-Gaussianity from the cosmic microwave
background \citep[e.g.,][]{Planck13NG} or higher order statistics
\citep[e.g.,][]{Marin13} to better calibrate the scale dependence of
the galaxy bias.

\section{Systematic errors due to the uncertainties in halo properties}\label{sec:sys}

\begin{figure*}
\includegraphics[width=\columnwidth]{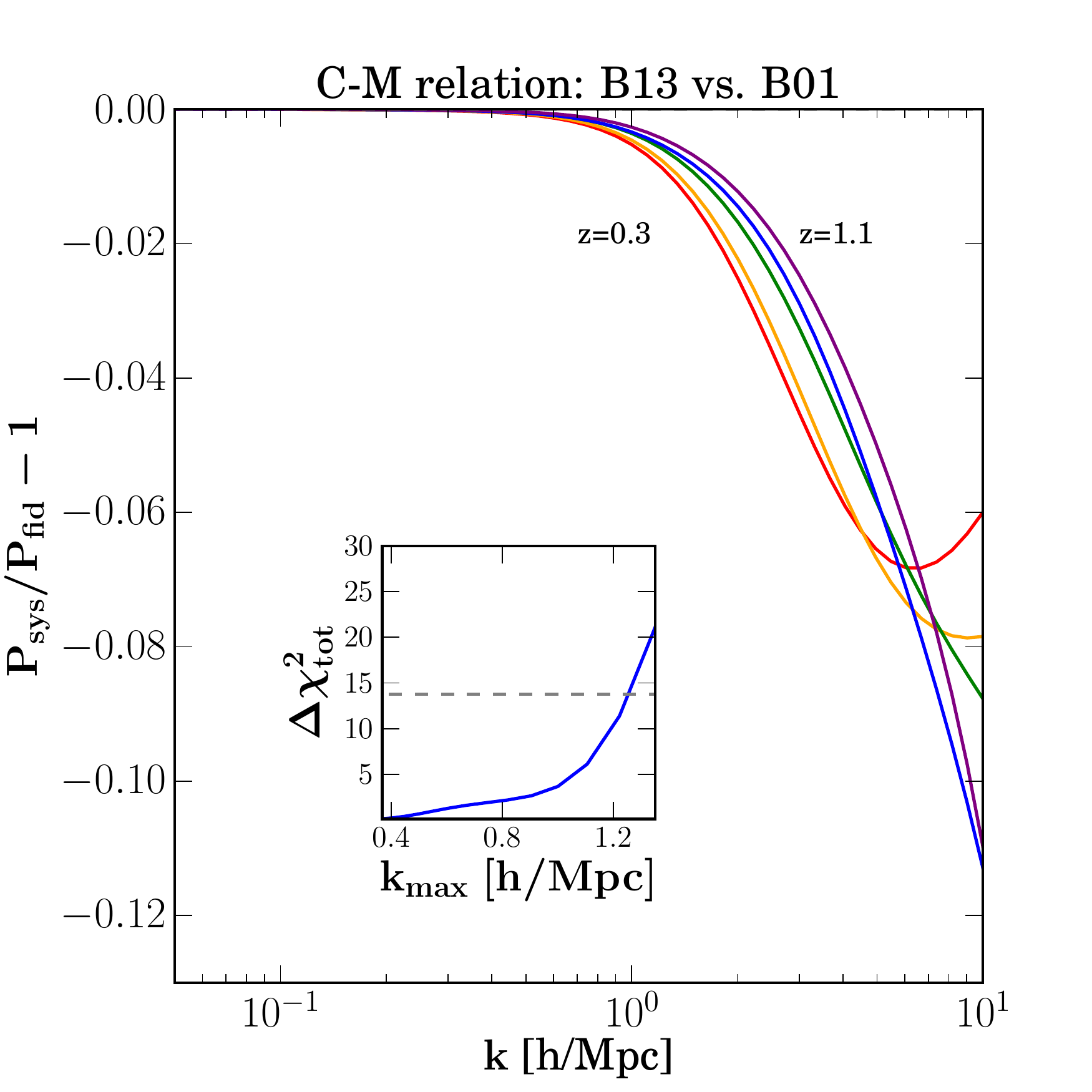}
\includegraphics[width=\columnwidth]{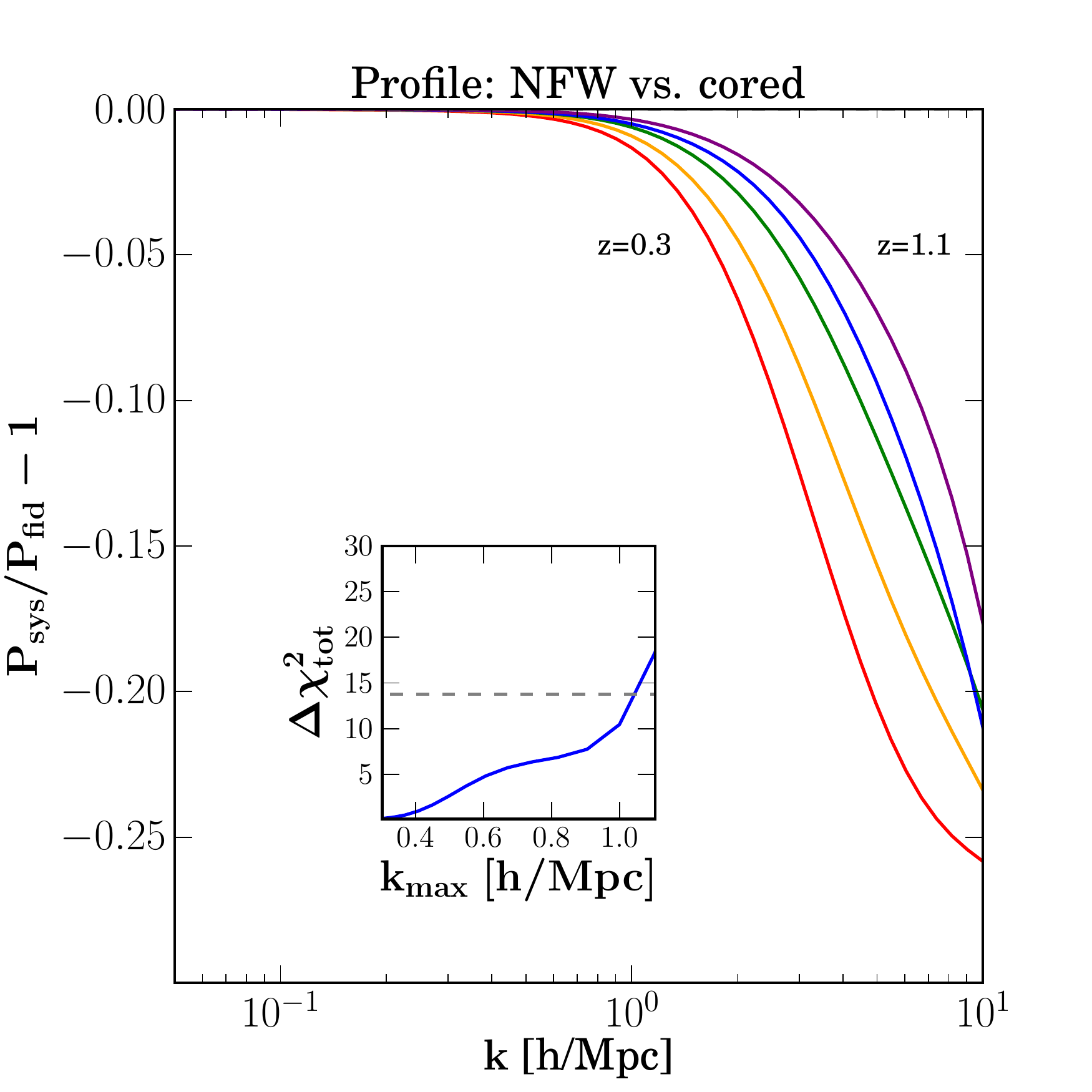}
\includegraphics[width=\columnwidth]{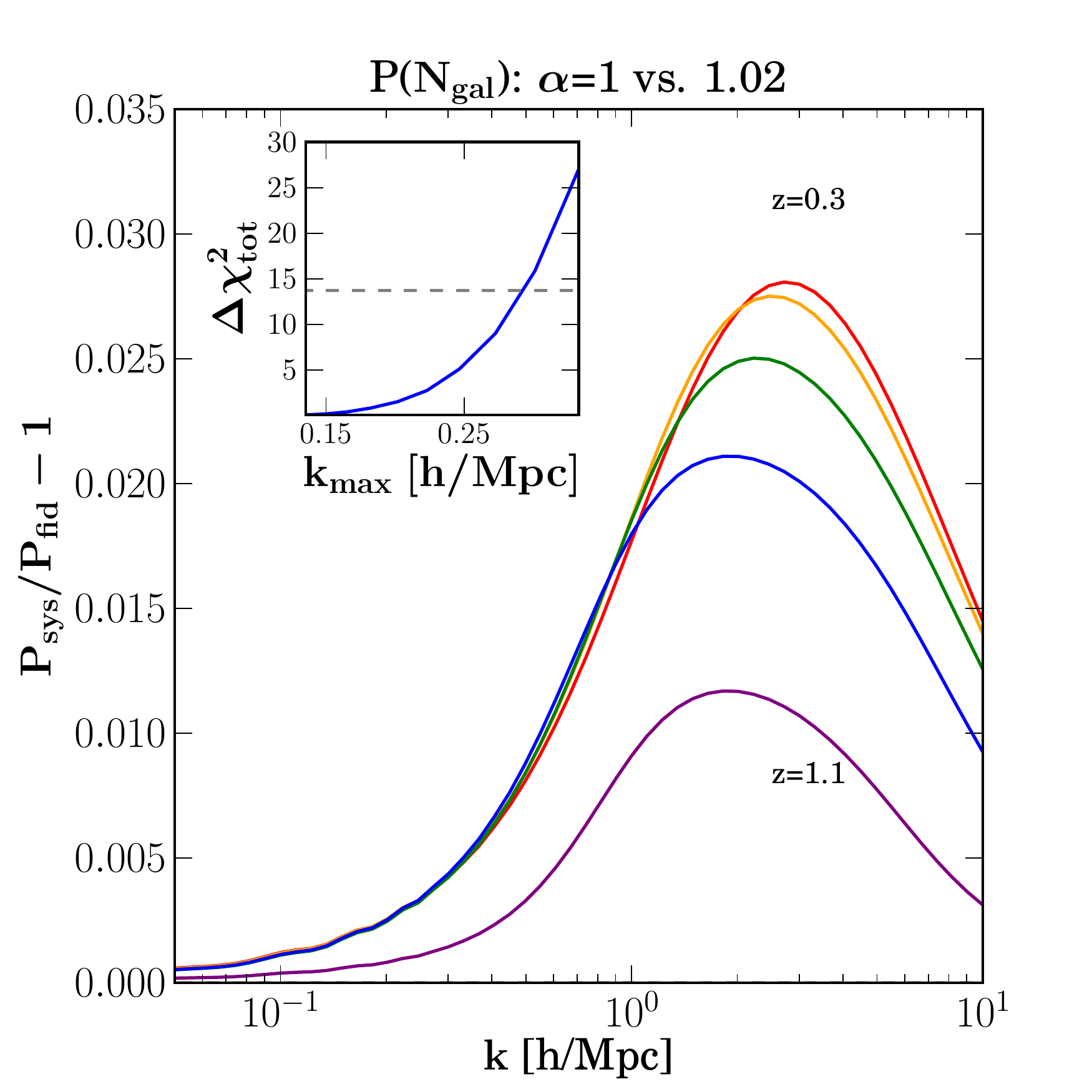}
\includegraphics[width=\columnwidth]{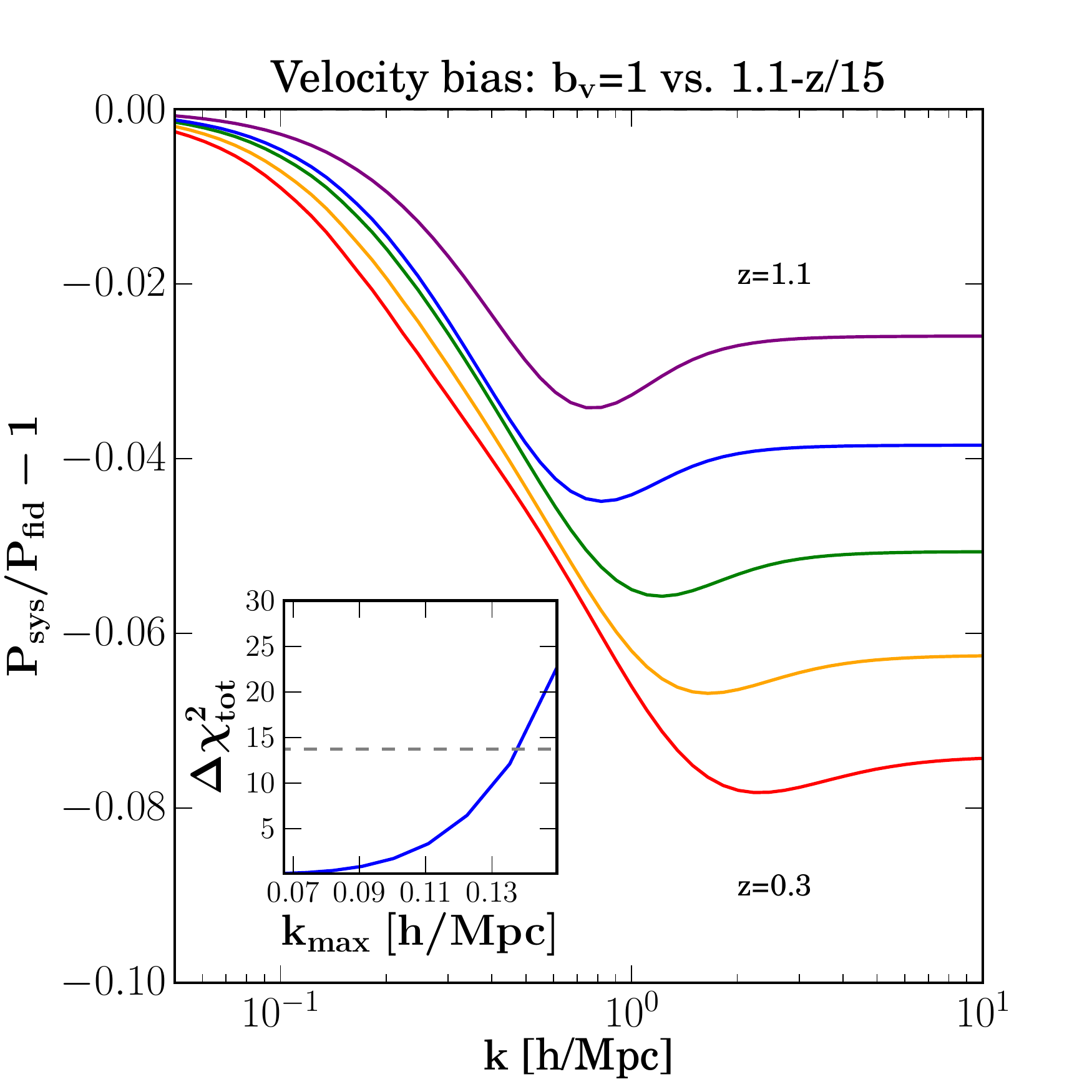}
\caption{Systematic differences in $P(k)$ caused by the four sources
of errors discussed in Section \ref{sec:sys}.  In each panel, the main
figure shows the fractional difference in $P(k)$ in the five redshift
bins, while the inset shows the systematic error $\Delta \chi^2_{\rm
tot}$ as a function of $\kmax$.  The 1 $\sigma$ deviation in the
12-dimensional parameter space, $\Delta\chi^2_{\rm tot} = 13.7$, is
marked by the horizontal dashed line in each inset.}
\label{fig:sys}
\end{figure*}

In this section, we explore the impact of four sources of theoretical
uncertainties related to the properties of dark matter haloes coming
from $N$-body simulations on the constraining power of $P(k)$.  These
sources of systematics are as follows:
\begin{itemize}
\item concentration--mass relation
\item deviation of $\tug$ from the NFW profile
\item deviation of $\Ns$ from the Poisson distribution
\item velocity bias  
\end{itemize}
In particular, we address the following points.
\begin{itemize}
\item With the current level of uncertainties, what are the systematic errors
  in the prediction of $P(k)$? What are the biases in the parameter inference caused by these systematics?
\item What is the smallest scale (largest $\kmax$) allowed by the current level of
  uncertainties?
\item What is the required reduction of these uncertainties if we would
  like to push to higher $\kmax$?
\end{itemize}
We again use  $\Delta\chi^2_{\rm tot}$ to assess the impact of
systematic errors on the cosmological parameters, as described in
the previous section.
The summary of the impact of these systematics is presented in
Fig.~\ref{fig:sys} and Table~\ref{tab:sys}.

\begin{table*}
\centering
\setlength{\tabcolsep}{0.5em} 
\begin{tabular}{|c|c|c||c|c|c||c|c|c|}
\hline
\rule[-2mm]{0mm}{6mm}
& Systematic & $\kmax$ & 
 \multicolumn{3}{c||}{$\kmax = 0.3\, \hoverMpc$} & \multicolumn{3}{c|}{$\kmax = 1\, \hoverMpc$} \\
\cline{4-9}
\rule[-2mm]{0mm}{6mm}
 & Difference & allowed & $\avg{\Delta P/P}$& Deviation $(\sigma)$ &  $\fsys$ req. & $\avg{\Delta P/P}$ & Deviation $(\sigma)$ &  $\fsys$ req. \\
\hline
\rule[-2mm]{0mm}{6mm} $c$--$M$ relation & B13 versus  B01 & 1.3 & 0.00019 & 0 & None & 0.0039 & 0.014 & None \\ \hline
\rule[-2mm]{0mm}{6mm} Profile & NFW versus cored & 1 & 0.0004 & 0 & None & 0.0074 & 0.56 & None \\ \hline
\rule[-2mm]{0mm}{6mm} P($\Ns$) & $\rm\alpha$=1 versus 1.02 & 0.29 & 0.0037 & 1.3 & 0.93 & 0.016 & 15 & 0.23 \\ \hline
\rule[-2mm]{0mm}{6mm} Velocity bias & $\rm b_v$=1 versus 1.1-z/15 & 0.14 & 0.026 & 32 & 0.11 & 0.052 & 108 & 0.034 \\ \hline
\end{tabular}
\caption{Summary of the effects of the four sources of systematic
error considered in Section \ref{sec:sys}. Note that $\avg{\Delta
P/P}$ is calculated at $k=\kmax$ and averaged over the five redshift
bins, and `$\fsys$ req.'  is the required reduction factor in the
amplitude of the systematic difference so that it becomes a 1 $\sigma$
effect in the full parameter space.}
\label{tab:sys}
\end{table*}

\subsection{Concentration--mass relation}\label{sec:c-M}

In the halo model, the one-halo term depends on the number density
profile of galaxies, $\tug(k|M)$. We assume that the galaxy
distribution follows the dark matter distribution, which is well
described by an NFW profile.  We then use the concentration--mass
relation of dark matter haloes from the literature to compute
$\tug(k|M)$.

The concentration--mass relation has been calibrated with dark matter
$N$-body simulations
\citep[e.g.,][]{Bullock01,Neto07,Duffy08,Maccio08,Kwan13,Prada11,Bhattacharya11}
and hydrodynamical simulations \citep[e.g.,][]{Lau09,Duffy10,Rasia13}.
Several observational programmes are also working towards pinning down
this relation \cite[e.g.,][]{Coe12,Oguri12}.  However, 10--20 per cent
of uncertainties in the concentration--mass relation remain, and the
concentration--mass relation also varies with cosmology and the
implementation of baryonic physics (see, e.g., the review in
\citealt{Bhattacharya11}).

We investigate the impact of uncertainties in the concentration--mass
relation by comparing the models from \citet[][B01
hereafter]{Bullock01} and the recent calibration from \citet[][B13
hereafter]{Bhattacharya11}.  These two models represent two extreme
cases of the concentration--mass relation; therefore, using these two
extreme cases sets the upper limit of the systematic bias caused by
the $c$--$M$ relation.  We assume a scatter of 0.33 for the $c$--$M$
relation in both cases.  Ignoring this scatter will lead to an
approximately 0.5 per cent difference in $P(k)$ at $k\approx 1\,
\hoverMpc$.

Our baseline model is from the recent formula given by B13 (based on virial
overdensity): 
\beqa
c(\nu) &=& D(z)^{0.78} 7.9 \nu^{-0.28}  \\[0.2cm]
\nu&=& \frac{1}{D(z)}\left[ 1.12 \left(\frac{\Mvir}{5\times 10^{13}\hiMsun}\right)^{0.3} + 0.53
\right]. \nonumber
\eeqa
We compare it with the model from B01:
\beq
c(\Mvir) = \frac{9}{1+z}\left(\frac{\Mvir}{M_*(z)}\right)^{-0.13}.
\eeq
These two calibrations agree near $M_*$ at $z=0$.

The top-left panel of Fig.~\ref{fig:sys} shows the relative change in
the power spectrum $P(k)$, evaluated at five redshifts, due to the
difference between B01 and B13.  We find that $P(k)$ based on B01 is
in general lower than that based on B13, because B01 predict lower
concentrations at the high-mass end.  Although B01 predict higher
concentrations at the low-mass end, these haloes rarely contribute to
the one-halo term and thus do not significantly boost clustering.

The inset in this panel shows the systematic shifts in the parameter
space caused by different models, which are characterized by
$\Delta\chi^2_{\rm tot}$.  It can be seen that the systematic error
starts to be comparable to the statistical error ($\Delta\chi^2_{\rm
tot} = 13.7$, marked by a horizontal dashed line) at $\kmax = 1.2 \
\hoverMpc$, which makes it a relatively unimportant source of
systematic error.

We would now like to study the effects of improved calibration in the
$c$--$M$ relation.  A natural way to do this is to assume that the
difference between the two extreme predictions has been reduced by
some constant factor, and that the new value interpolates between the
two original extremes.  We define the interpolated value as
\beq
c_{\rm interp}(M) = c_{\rm fid}(M) + \fsys \left(c_{\rm alt}(M) - c_{\rm fid}(M)\right) \ ,
\eeq
where $c_{\rm fid}(M)$ and $c_{\rm alt}(M)$ are respectively the
fiducial (say, B13) and the alternate (say, B01) models for the
concentration--mass relation.  Here $\fsys$ is a tunable parameter that
allows us to assess the effect of a fraction of the full
systematics. The limiting cases are:
\begin{align*}
\fsys = 0 &\quad\Longleftrightarrow \quad\mbox{no systematics}\\[0.1cm]
\fsys = 1 &\quad\Longleftrightarrow \quad\mbox{fiducial systematics \ .}
\end{align*}
For a higher $\kmax$, the tolerance of systematics is smaller, and
$\fsys$ provides a measure for required reduction of systematics.  For
a given $\kmax$, we search for the appropriate $\fsys$ value that
makes the systematic negligible\footnote{Note that some fraction
$\fsys$ of the systematics does {\it not} trivially lead to the same
fractional shift in $P(k)$ because the $c$--$M$ relation (and most
other systematics) enters non-linearly into $P(k)$.  We therefore need
to perform a separate calculation of $P(k)$ for each $\fsys$.}.

\begin{figure}
\includegraphics[width=\columnwidth]{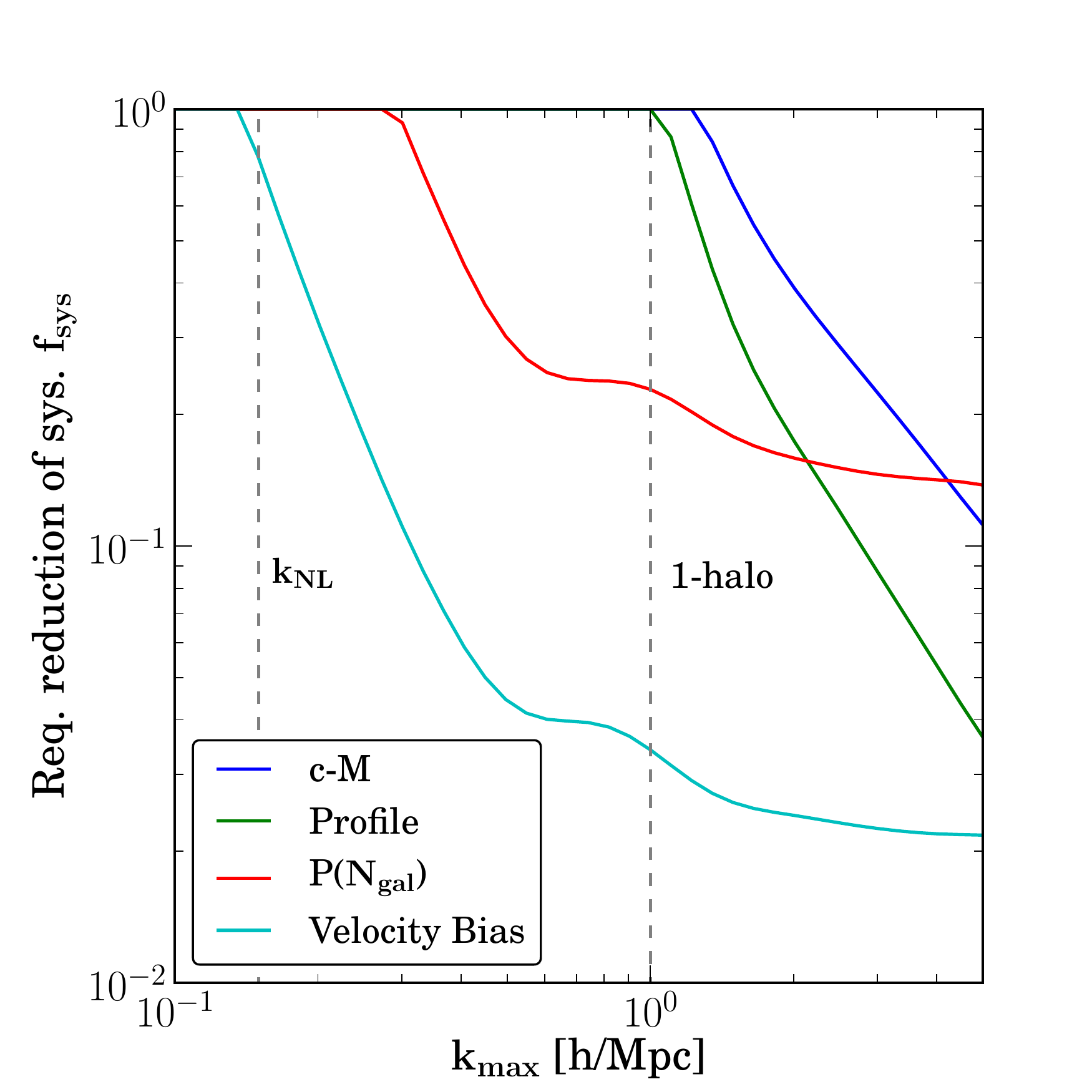}
\caption{Required reduction of systematic errors, shown as the
fraction of the current errors, for the four sources of systematic
errors discussed in this paper, as a function of the maximum
wavenumber considered in the survey.  Note that the velocity bias
requires the greatest improvement relative to the current knowledge.}
\label{fig:freq}
\end{figure}

The blue curve in Fig.~\ref{fig:freq} shows the requirement on $\fsys$
from the c-M relation as a function of $\kmax$.  For all practical
$\kmax$ values, $c$--$M$ does not require more precise calibrations
from $N$-body simulations.  The results are summarized in the
`$c$--$M$ relation' row of Table~\ref{tab:sys}.

\subsection{Galaxy number density profile: deviation from NFW}

Our fiducial model assumes that the galaxy distribution inside a halo
is described by the NFW profile.  However, $N$-body simulations have
shown that the distribution of subhaloes in cluster-size haloes tends
to be shallower than the NFW profile, and also shallower than the
observed galaxy number density profile
\citep[e.g.,][]{Diemand04,NagaiKravtsov05}.  These deviations could be
related to insufficient resolution or the absence of baryons in
$N$-body simulations -- the so-called overmerging issue.  Several
authors have proposed models for `orphan galaxies' to compensate the
overmerging issue; however, these models do not always recover the
observed galaxy clustering \cite[e.g.,][]{Guo11}.  The exact cause for
these issues is still uncertain; nevertheless, the uncertainties
associated with the distribution of subhaloes will likely impact the
modelling of galaxy clustering.  Based on the comparisons between dark
matter and hydrodynamical simulations
\citep[e.g.,][]{Maccio06,Weinberg08}, the observed galaxy density
profile is likely to be bracketed by the density profiles of the
subhalo number and dark matter.

In this section, we investigate whether the uncertainties in the
galaxy number density profile lead to a significant systematic bias.
To model the possibility that the galaxy distribution is shallower
than dark matter in the inner region of clusters, we adopt the subhalo
number density profile measured from Wu et al.~(in preparation), which
is also illustrated in Appendix~\ref{app:Consuelo}.
Fig.~\ref{fig:Consuelo} presents one example of the galaxy number
density profile measured from an $N$-body simulation.  Based on this
result, we model the subhalo number density profile as
\beq
u(r | M) = f_{\rm surv}(M,r) \times u_{\rm NFW}(r|M)  \ ,
\label{eq:orphan_profile}
\eeq
where $u_{\rm NFW}(r|M)$ is the  NFW profile , and $ f_{\rm surv}$ is 
the ``surviving fraction'' of galaxies given by
\begin{equation}
\begin{aligned}
f_{\rm surv} &= 1-0.99 {\rm e}^{-a (r/\Rvir)} \\ 
a &\equiv  0.005 \left( \ln \frac{\Mvir}{1000\hiMsun} \right)^2  \ .
\end{aligned}
\end{equation}
We note that $f_{\rm surv}$ is smaller for higher host halo mass and
smaller radius, where the effect of overmerging is stronger.

The top-right panel of Fig.~\ref{fig:sys} shows the difference in
$P(k)$ caused by this cored profile.  As expected, the deficit of the
galaxy number at small scales leads to lower power at high $k$.  In
addition, the suppression is stronger at low redshift because massive
clusters are more abundant at low $z$.  The inset shows the
corresponding $\Delta\chi^2_{\rm tot}$ as a function of $\kmax$; the
systematic shifts dominate at $\kmax = 1\, \hoverMpc$.

We model the interpolated systematic error in the density profile as
\beq
\tug_{\rm interp}(k|M) = \tug_{\rm fid}(k|M) + \fsys \left[\tug_{\rm alt} (k|M)- \tug_{\rm fid}(k|M)\right]  \ .
\eeq
Here our fiducial model is the NFW profile, and the alternative
profile is given by equation~\ref{eq:orphan_profile}.  We again search
for the required $\fsys$ as a function of $\kmax$.  The result is
shown by the green curve in Fig.~\ref{fig:freq}.  Like the $c$--$M$
relation, the density profile of galaxies does not require more
precise calibrations for all practical $\kmax$.  The results are
summarized in the `Profile' row of Table~\ref{tab:sys}.

\subsection{Deviation from the Poisson distribution}

In our fiducial model, $P(\Ns|M)$ is assumed to be Poisson
distributed; that is, the second moment is given by $\avg{\Ns(\Ns-1)}
= \avg{\Ns}^2$, or $\alpha\equiv {\sqrt{\avg{\Ns(\Ns-1)}}}/{\avg{\Ns}}
= 1$.  However, \citet{Boylan-Kolchin10} have shown that the number of
subhaloes for a given halo mass deviates from the Poisson distribution
(their fig.  8).  In addition, \citeauthor{Wu12b} (2013a) have shown
that the extra-Poisson scatter depends on how subhaloes are chosen and
depends on the resolution.  Therefore, it is still unclear whether
$P(\Ns|M)$ follows a Poisson distribution.  To assess the impact of
the possible extra-Poisson scatter, we adopt $\alpha = 1.02$ in our
one-halo term (following \citealt{Boylan-Kolchin10}), noting that this
choice of $\alpha$ brackets the various possibilities explored in
\citeauthor{Wu12b} (2013a, fig.~3 therein).

The bottom-left panel in Fig.~\ref{fig:sys} shows the impact of
$\alpha=1.02$ on $P(k)$, relative to the fiducial Poisson case with
$\alpha=1$.  The extra-Poisson scatter only impacts the one-halo term;
therefore, the large-scale $P(k)$ is unaffected.  At small scales,
$P(k)$ is boosted by less than 3 per cent.  For different redshifts,
$\Delta P/P$ takes off at different $k$, reflecting the varying scale
where one-halo and two-halo terms cross.  We also note that at high
$k$, $\Delta P/P$ bends downwards, reflecting the fact that the
one-halo term includes $\avg{\Ns|M}\tug + \frac{1}{2}
\avg{\Ns(\Ns-1)|M} \tug^2$.  When $P(\Ns|M)$ is super-Poisson, more
galaxy pairs are expected, and the one-halo term gets more weighting
of $\tu^2$ ($\tu<1$); thus, $P(k)$ becomes lower at high $k$.

The inset in the bottom-left panel of Fig.~\ref{fig:sys} shows that
the systematic shifts dominate statistical errors at $\kmax=0.3\
\hoverMpc$.  We model the partial uncertainties in $\alpha$ as
\beq
\alpha_{\rm interp} = \alpha_{\rm fid} + \fsys (\alpha_{\rm alt} - \alpha_{\rm fid}) \ .
\eeq
The red curve in Fig.~\ref{fig:freq} shows the required $\fsys$ as a
function of $\kmax$; for $\kmax=1\, \hoverMpc$, the required $\fsys =
0.2$.  The results are summarized in the $P(\Ns)$ row of
Table~\ref{tab:sys}.

\subsection{Velocity bias}

The small-scale RSD (also known as the `Fingers-of-God' effect) are
usually modelled as an exponential suppression of power with the term
${\rm e}^{-(k\sigmav\mu)^2}$.  Here $\sigmav$ is the velocity
dispersion of {\it galaxies} inside a cluster $\sigmav^{\rm
gal}(\Mvir)$.  Assuming that the motions of galaxies trace those of
dark matter particles, we use the velocity dispersion of dark matter
particles inside a halo, $\sigmav^{\rm DM}(\Mvir)$, which has been
well established using simulations \citep{Evrard08}.  However, the
velocity dispersion of {\it galaxies} inside a cluster $\sigmav^{\rm
gal}$ is not necessarily the same as $\sigmav^{\rm DM}$.  The ratio
between the two is defined as the velocity bias
\beq 
\bv = \frac{\sigmav^{\rm gal}}{\sigmav^{\rm DM}} \ .
\eeq 
The exact value of $\bv$ and its redshift dependence are still under
debate.  Subhaloes from $N$-body simulations have shown $\bv > 1$
\citep[e.g.,][]{Colin00}.  In addition, \citeauthor{Wu13b} (2013b)
have shown that the exact value of $\bv$ depends on the selection
criteria applied to subhaloes, on the resolution of simulations, and
on the location of subhaloes.  On the other hand, a simulated galaxy
population based on assigning subhaloes to dark matter particles
\citep[e.g.,][]{Faltenbacher06} or based on hydrodynamical simulations
with cooling and star formation \citep[e.g.,][]{Lau10,Munari13} tends
to have unbiased velocities.

Since this paper focuses on the possible systematics from $N$-body
simulations, we adopt $\bv>1$ observed in $N$-body simulations.  Based
on the recent calibration from \cite{Munari13}, we adopt the value of
velocity bias to be
\beq
\bv(z) = 1.1 - \frac{z}{15}
\eeq
(estimated from the dotted curve in their fig.\ 7A, which corresponds
to subhaloes in their $N$-body simulations.)  The bottom-right panel
of Fig.~\ref{fig:sys} shows the systematic error in $P(k)$ caused by
this velocity bias.  Introducing higher velocity dispersion of
galaxies clearly leads to larger suppression on small scales.  Note
that each curve showcases a dip near $k\approx1\, \hoverMpc$, which
roughly corresponds to the scale where one-halo and two-halo terms
cross.  As shown in Section \ref{sec:Pk_rsd}, the exponential
suppression of RSD enters the one-halo and two-halo terms differently;
modifying the RSD will therefore slightly change the scale of one-halo
to two-halo transition.  Also note that the shift in $P(k)$ does not
vanish even for very small $k$, because the exponential suppression
enters the two-halo term as well.

The inset in the bottom-right panel of Fig.~\ref{fig:sys} shows that
the systematic shifts associated with velocity bias (difference
between no velocity bias and positive velocity bias) dominate the
statistical error even for $\kmax = 0.14 \, \hoverMpc$.  Because the
deviation of $P(k)$ starts at large scales and increases towards small
scales, the velocity bias is dominant among the four sources of
systematic errors studied in this paper.

As before, we consider values of the velocity bias that interpolate
between the two extreme values considered:
\beq
{\bv,\,}_{\rm interp} = {\bv,\,}_{\rm fid} + \fsys \left({\bv,\,}_{\rm alt} - {\bv,\,}_{\rm fid} \right) \ ,
\eeq
where $\fsys=0$ corresponds to $\bv={b}_{\rm fid} =1.0$ while
$\fsys=1.0$ corresponds to $\bv={\bv,\,}_{\rm alt}=1.1-z/15$.  The
cyan curve in Fig.~\ref{fig:freq} shows the required reduction of
$\fsys$ for a given $\kmax$.  For example, to extend the survey just
out to the usually conservative wavenumber $\kmax = 0.3\, \hoverMpc$,
better-than-current knowledge of the velocity bias ($\fsys=0.11<1$) is
required.\footnote{Note that \citet{Colin00} have shown that $\bv$ is
scale dependent.  Since our scale independent assumption has already
introduced significant systematic shifts, we do not further consider
the possible scale dependence of velocity bias in this work but note
that the possible scale-dependence will further complicate the
systematic error.}

Given that a biased $\bv$ value can lead to a significant systematic
error, it is necessary to marginalize over $\bv$ to mitigate the
systematic bias.  We find that marginalizing over an additional
parameter $\bv$ in the Fisher matrix calculation does not
significantly degrade the dark energy constraints; the statistical
error $\sqrt{\sigma(w_a)\sigma(w_p)}$ is increased by a factor of 2 at
most.  Since $P(k)$ is sensitive to the change in $\bv$ (as shown in
the last panel of Fig.~\ref{fig:sys}), it is not surprising that $\bv$
can be well constrained by data when set free.  In addition, the
effect of $\bv$ does not seem to be degenerate with the effects of
other nuisance parameters and is likely to be well constrained.

While the preparation of this paper was near completion, we learned
about the related work from \citet{LinderSamsing13}. These authors
have focused on a particular RSD model from \cite{Kwan12} and assess
the impact of uncertainties in this model on cosmological constraints.
These authors have found that, if the model parameters are fixed, they
often require sub-per cent accuracy; on the other hand, if these model
parameters are self-calibrated using the data, they do not
significantly degrade the cosmological constraints.  This trend is
consistent with our findings regarding fixing versus marginalizing
over the velocity bias.

We emphasize that the main goal of this paper is to see to what extent
the theoretical uncertainties associated with calibrating galaxy
clustering using $N$-body simulations lead to errors in the
cosmological parameters.  Given the difficulty of predicting
clustering beyond $k\simeq 0.5\, \hoverMpc$ using purely theoretical
methods (e.g., the perturbation theory), resorting to calibration with
$N$-body simulations is required, and this will remain to be the case
for years to come.  Our findings suggest that the velocity information
of galaxies predicted from $N$-body simulations is likely to generate
biases.

\section{Summary}\label{sec:summary}

As the interpretation of the galaxy clustering measurements from deep,
wide redshift surveys often relies on synthetic galaxy catalogues from
$N$-body simulations, the systematic uncertainties in $N$-body
simulations are likely to lead to systematic errors in the
cosmological results.  In this paper, we have studied several
theoretical uncertainties in the predictions of $N$-body simulations,
including the statistics, the spatial distribution and the velocity
dispersion of subhaloes.  In particular, we have applied the halo
model to calculate the galaxy power spectrum $P(k)$, with inputs from
recent $N$-body simulations.  We have investigated how the
uncertainties from these inputs impact the cosmological interpretation
of $P(k)$, and how well these systematics need to be controlled for
future surveys.  Our main findings can be summarized as follows:

\begin{itemize}

\item We have found that the inclusion of the RSD and the covariances
between different $k$ modes (the trispectrum contribution to the
covariance matrix) is essential to accurately model the information
content at small scale.

\item Uncertainties in the halo mass function and bias tend to affect
$P(k)$ on large scales and can lead to significant systematic errors.
However, these effects can be mitigated by measurements of galaxy bias
at large scales combined with an independent measurement of
$\sigma_8$.

\item Uncertainties in predicting the halo concentration--mass
relation, as well as the deviation from an NFW profile, are unlikely
to be a dominant source of systematic error for $\kmax<1 \,
\hoverMpc$.

\item Possible deviation of $P(\Ns)$ from the Poisson distribution, at
its current uncertainty level (2 per cent) could be significant for
$\kmax > 0.3 \, \hoverMpc$.

\item Velocity bias is likely to be the most important source of
systematic error for upcoming surveys.  The current uncertainty of 10
per cent at $z=0$ is likely to introduce 3 (5) per cent difference in
$P(k)$ for $\kmax$ = 0.3 (1) $\hoverMpc$, thus leading to a
significant bias in cosmological parameters.  Given its predominant
role in the systematics, the velocity bias will need to be calibrated
internally from the survey or externally with follow-up campaigns.

\end{itemize}

The sensitivity of $P(k)$ to velocity bias leads to the question of
what can be done to alleviate the potential systematic
bias. Calibration through both observations and simulations is
certainly one obvious solution. Another trick that is increasingly
being used for large-scale structure surveys is to self-calibrate the
systematic error(s); in the velocity-bias case, this would mean
marginalizing over $\bv$.  With this marginalization, we expect to be
left with vastly diminished biases and only a modest degradation in
the cosmological parameters. We do not expect the bias to vanish
completely, however, since second-order effects (e.g., redshift- and
scale-dependence of $\bv$) will remain and will cause systematic
shifts. Given that we currently do not have a good model of $\bv(z,
k)$, we have not attempted the full self-calibration exercise, but we
definitely expect this to be modus operandi of galaxy clustering
analyses in the future.

\section*{Acknowledgements}
We thank Andrew Hearin, Eric Linder, Chris Miller, and Zheng Zheng for many helpful suggestions.
We also thank the anonymous referee for helpful comments.
This work was supported by the U.S.\ Department of
Energy under contract number DE-FG02-95ER40899.

\bibliographystyle{mn2e}
\bibliography{master_refs}

\appendix

\section{Galaxy number density profile}\label{app:Consuelo}
\begin{figure}
\includegraphics[width=\columnwidth]{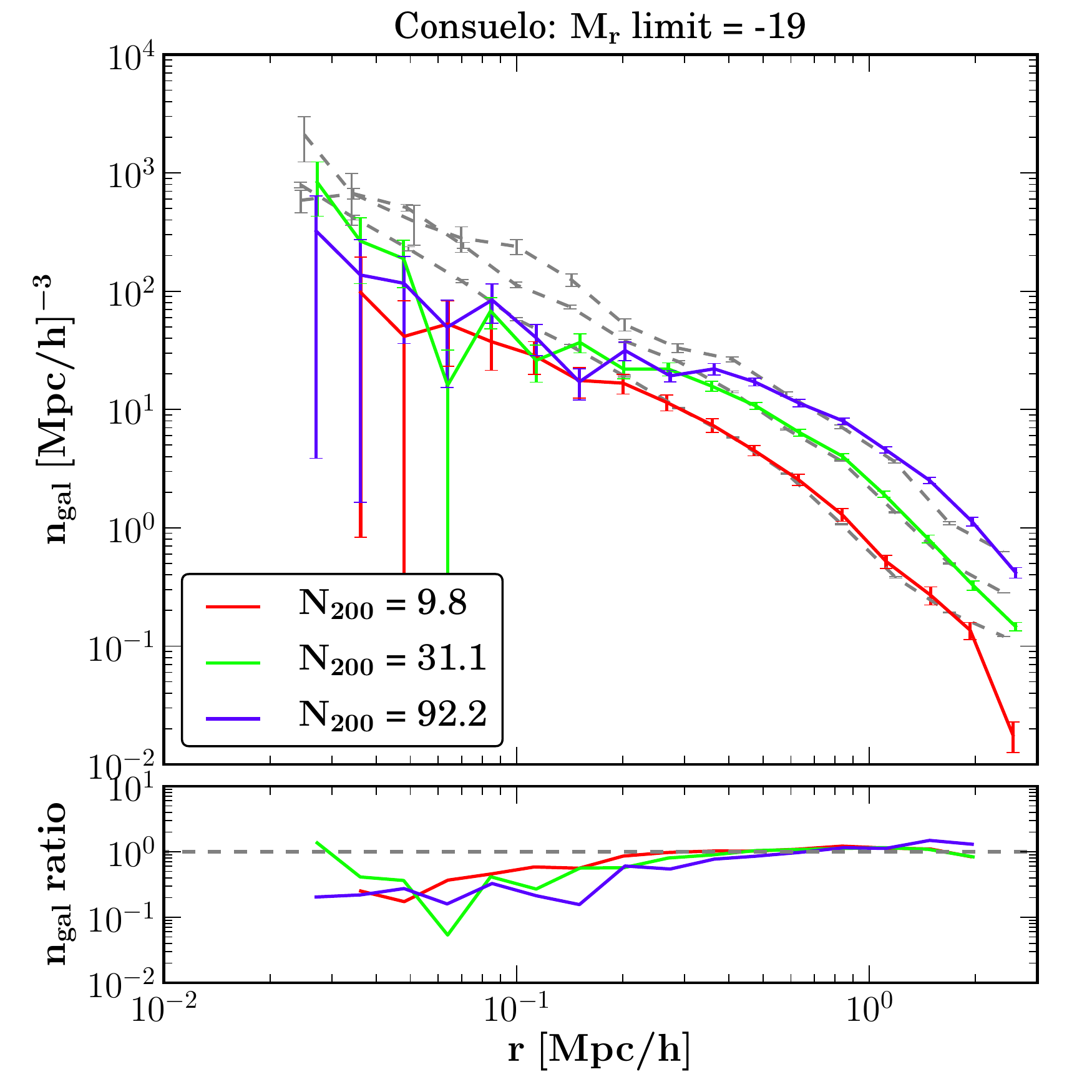}
\caption{Galaxy number density profile from the Consuelo simulation.
The colored curves show galaxies in the simulation, while the grey
dashed curves correspond to the observations of SDSS from
\protect\cite{Tinker12}.  As can be seen, the simulated galaxy
population near the centre of clusters tend to have a shallower
distribution than the real galaxy population.  The difference is
larger for more massive clusters.  This figure is adapted from Wu et
al.~(in preparation).}
\label{fig:Consuelo}
\end{figure}

Fig.~\ref{fig:Consuelo} presents the galaxy number density profile
based on which we model its theoretical uncertainties.  The colored
curves are based on the Consuelo simulation -- an $N$-body simulation
with $1400^3$ particles in a volume of side length $420\, \hiMsun$.
The mass resolution is $1.9\times10^9\, \hiMsun$, and the force
resolution is $8\, \hikpc$.  We assign each subhalo a luminosity value
using the $v_{\rm max}^{\rm pk}$--luminosity relation based on a
subhalo abundance matching model (Behroozi, private communication),
where $v_{\rm max}^{\rm pk}$ is the subhalo's peak maximum circular
velocity in its history.

We compare the simulated galaxy density profiles with the results from
the SDSS maxBCG cluster catalogue as presented in \citet{Tinker12}.
The grey dashed curves correspond to three of the richness bins of
maxBCG.  From the Consuelo simulation, we select clusters in a way
that they have approximately the same mass distribution as the maxBCG
cluster sample \citep{Johnston07}.  Each maxBCG cluster is assigned a
richness value $N_{200}$, which is the number of red-sequence galaxies
brighter than $^{0.25}M_i = -19.2$ within $r_{200}$. Here $r_{200}$ is
defined as the radius within which the density of galaxies is 200
times the mean density of galaxies.  At large radii ($> 0.5\,
\hiMpc$), the simulation and observation agree well. This agreement
naturally comes from our mass selection and abundance matching without
tuning the normalization.

However, discrepancy between simulation and observation occurs at
small radius.  As can be seen, the subhalo number density profile
measured from the simulation is shallower than the galaxy density
profile measured from SDSS and is also shallower than the NFW profile.
This discrepancy is stronger for more massive host haloes.  Wu et
al.~(in preparation) further demonstrate that (1) the discrepancy is
also stronger for dimmer galaxies, (2) the trend exists in several
state-of-the-art $N$-body simulations using different algorithms and
resolutions, and (3) the incompleteness of subhaloes depends on the
radius, the mass of the host halo and the mass of the subhalo.  It has
been shown that the deficit of simulated galaxies near the centre of
massive haloes can be alleviated in hydrodynamical simulations that
include cooling and star formation
\citep[e.g.,][]{Weinberg08,Dolag09}.  Therefore, it is highly likely
that this deficit presents a fundamental limitation of $N$-body
simulations and needs to be taken into account when we use $N$-body
simulations to model the galaxy population in massive clusters.

\section{Derivation of the galaxy power spectrum}\label{app:halomodel}

In this appendix, we provide the detailed derivation of the galaxy
power spectrum, mainly following the derivations in
\cite{ScherrerBertschinger91}, \cite{Seljak00} and
\cite{CooraySheth02}, in order to clarify possible confusions
originated from different conventions.  
Let us assume that dark matter halo $i$ with mass $M_i$ is located at
$\vx_i$.  It has $N_i$ galaxies, whose spatial distribution is
described by $u(\vx-\vx_i|M_i)$ [normalized so that $\int {\rm d}^3\vx\
u(\vx| M) = 1$].  The galaxy number density field can be described by
summing over all haloes in the universe:
\beq\bal
&\rhog(\vx) = \sum_i N_i \ u(\vx -\vx_i | M_i) \\
&= \sum_i \int dM \deltaD(M-M_i) \int {\rm d}^3\vx' \deltaD(\vx' - \vx_i)  N_i u\left(\vx - \vx' |M \right),
\eal\eeq
where we insert Dirac delta functions for $M$ and $\vx'$.
If we define
\beq
\begin{aligned}
\left\langle 
\sum_i  \deltaD(M-M_i)  \deltaD(\vx' - \vx_i) \ N_i
  \right\rangle \equiv n(M) \avg{N|M}  \ , 
\end{aligned}
\eeq
then the mean galaxy number density is given by
\beq
\begin{aligned}
{\brhog}= \avg{\rhog}= \int dM n(M)  \avg{N|M}  \ ,
\end{aligned}
\eeq
where we write $n(M)$ = ${dn}/{dM}$ for the halo mass function.

The number density fluctuation of galaxies is defined as
\beq
\deltag(\vx) = \frac{\rhog(\vx)}{\brhog} - 1 \ .
\eeq
The two-point statistics follows the definition:
\beq\bal
&\left\langle  
\sum_i \deltaD(M_1-M_i) \deltaD(\vx_1-\vx_i)  N_i
\sum_j \deltaD(M_2-M_j) \deltaD(\vx_2-\vx_j)  N_j
\right\rangle \\[0.2cm]
&\equiv  n(M_1) \avg{N|M_1}n(M_2) \avg{N|M_2}  \left[1+\xi_{\rm hh}(M_1, M_2, |\vx_2-\vx_1|) \right]  \quad (i\neq j) \\[0.2cm]
&+n(M_1)\avg{{N\choose2}|M_1}\deltaD(M_1-M_2)\deltaD(\vx_1-\vx_2)   \quad (i = j)  \ ,
\label{eq:2pt_real}  
\eal\eeq
where $\xi_{\rm hh}$ is the two-point correlation function of dark matter contributed by two different haloes.
The two-point correlation function for galaxies reads
\beq\bal
&\xi_{\rm gg}(r) = \avg{\deltag(\vx)\deltag(\vx+\vecr)} \\[0.2cm]
&= \frac{1}{\brhog^2} \int dM_1\int dM_2 \int {\rm d}^3\vx_1 \int {\rm d}^3\vx_2\  u(\vx-\vx_1|M_1)  u(\vx+\vecr-\vx_2|M_2)  \\
&\left\langle  \sum_i \deltaD(M_1-M_i) \deltaD(\vx_1-\vx_i)  N_i
\sum_j \deltaD(M_2-M_j) \deltaD(\vx_2-\vx_j)   N_j \right\rangle \\[0.2cm]
&= \xi_{\rm gg}^{\rm 1h}(r) + \xi_{\rm gg}^{\rm 2h}(r) \ , 
\eal\eeq
where
\beq\bal
&\xi_{\rm gg}^{\rm 1h} (r) = \frac{1}{\brhog^2} \int dM n(M) \avg{\textstyle{N\choose2}|M} \int {\rm d}^3\vx\ u(\vx|M) u(\vx+\vecr|M) \ ,\\[0.2cm]
&\xi_{\rm gg}^{\rm 2h} (r) = \frac{1}{\brhog^2} \int dM_1 n(M_1)\avg{N|M_1}  \int dM_2 n(M_2)  \avg{N|M_2} \\
&\int {\rm d}^3\vx_1 \int {\rm d}^3\vx_2 u(\vx_1|M_1) u(\vx_2-\vecr|M_2) (1+\xi_{\rm hh}(M_1,M_2; |\vx_2-\vx_1-\vecr|)) \ .
\eal\eeq
We now turn to the Fourier space.  
We follow this convention of the Fourier transform
\beq
\tilde\delta(\vk) = \frac{1}{\sqrt{V}} \int  {\rm d}^3\vx\  \delta(\vx)\  {\rm e}^{-i \vk \cdot \vx} \ .
\eeq
The Dirac delta function in k-space is defined as
\beq
\deltaD(\vk) = \frac{1}{V}\int \frac{{\rm d}^3\vx}{(2{\rm\pi})^3}\ {\rm e}^{-i\vk\cdot \vx} \quad \mbox{(dimensionless)} \ .
\eeq
From this convention, the relation between the correlation function and the power spectrum follows:
\beq
\xi(r) = \frac{1}{(2{\rm\pi})^3} \int {\rm d}^3\vk P(\vk) {\rm e}^{-i \vk\cdot\vecr} \ .
\eeq
The Fourier transform of the density perturbation reads
\beq\bal
\tdeltag(\vk) &= \frac{1}{\sqrt{V}} \int  {\rm d}^3\vx\  \deltag(\vx)\  {\rm e}^{-i \vk \cdot \vx} \\
&= \frac{1}{\brhog\sqrt{V}} \sum_i N_i \tug(\vk|M_i) {\rm e}^{-i \vk\cdot\vx_i} - {(2 {\rm\pi})^3}{\sqrt{V}}\deltaD(\vk) \ ,
\label{eq:delta_g}
\eal\eeq
where
\beq
\tug(\vk|M) = \int {\rm d}^3\vx u(\vx|M) {\rm e}^{-i\vk\cdot\vx}  \ .
\eeq
Based on this definition, $\tug\rightarrow 1$ when $k \rightarrow 0$ and is dimensionless.
Applying the Fourier transform to equation~(\ref{eq:2pt_real}), we obtain
\beq\bal
\label{eq:2pt_k}
&\left\langle  
\sum_i \deltaD(M_1-M_i) {\rm e}^{-i\vk_1\cdot\vx_i} N_i \sum_j \deltaD(M_2-M_j) {\rm e}^{+i\vk_2\cdot\vx_j} N_j
\right\rangle \\[0.2cm]
&\equiv n(M_1)\avg{N|M_1}n(M_2)\avg{N|M_2}(2{\rm\pi})^6 V^2 \deltaD(\vk_1)\deltaD(\vk_2)  \\[0.2cm]
&+(2{\rm\pi})^3 V n(M_1)\avg{N|M_1}n(M_2)\avg{N|M_2} P_{\rm hh}(M_1,M_2;k)\deltaD(\vk_1-\vk_2)\\[0.2cm]  
&+(2{\rm\pi})^3 V n(M_1) \avg{\textstyle{N\choose2}|M_1}\deltaD(M_1-M_2)\deltaD(\vk_1-\vk_2)
\eal\eeq
%
We note that under our convention of the Fourier transform, $\int {\rm d}^3\vk \delta_D(\vk) = 1/V$  and $\delta_D(\vk)\delta_D(\vk) = \delta_D(\vk)/(2{\rm\pi})^3$.

We are now ready to compute the  galaxy power spectrum. Applying the trick of inserting Dirac delta functions and then using equation~(\ref{eq:2pt_k}),  we obtain
\beq\bal
&P_{\rm gg}(k) = \frac{1}{(2{\rm\pi})^3}\avg{\tdeltag(\vk) \tdeltag^*(\vk)} \\
&=  \frac{1}{(2{\rm\pi})^3 V {\brhog}^2}
 \left\langle
\sum_i \tug(\vk|M_i) {\rm e}^{-i \vk\cdot\vx_i}   N_i
\sum_j \tug^*(\vk|M_j) {\rm e}^{i \vk\cdot\vx_j} N_j
\right\rangle  \\[0.2cm]
&=  P_{\rm gg}^{\rm 1h}(k) + P_{\rm gg}^{\rm 2h}(k) \ ,
\eal\eeq
where the two-halo term reads
\beq
P_{\rm gg}^{\rm 2h}(k)= 
\left[\frac{1}{{\brhog}}
\int dM \frac{dn}{dM} \avg{N|M} b(M)
\right]^2 P_{\rm lin}(k) \ , 
\eeq
and the one-halo term reads
\beq
P_{\rm gg}^{\rm 1h}(k)=\frac{1}{{\brhog}^2} \int dM \frac{dn}{dM}\avg{\displaystyle{\left . N\choose 2\right|}M}f(k|M)  \ . 
\eeq
Here $\avg{\textstyle{\left . N\choose 2\right|}M}f(k|M)$ is the galaxy pair-weighted profile, 
including the contribution from 
central and satellite galaxies \citep{BerlindWeinberg02} 
\beq\bal
&\avg{\displaystyle{\left . N\choose 2\right|}M}f(k|M) \\
&= \left[
\avg{\Ns|M}\tug(k|M) + \frac{1}{2} \avg{\Ns(\Ns-1)|M}  |\tug(k|M)|^2 
\right]  \ .
\eal\eeq


\section{Derivation of the covariance matrix}\label{app:Cij}

We now derive the covariance of power spectra at different wave numbers in equation~(\ref{eq:C_ij}).
First, recall the definitions for power spectrum and trispectrum:
\beq\bal
\avg{\delta(\vk_1) \delta(\vk_2)} &= (2{\rm\pi})^3 \deltaD(\vk_{12}) P(k_1) \\[0.2cm]
\avg{\delta(\vk_1) \delta(\vk_2) \delta(\vk_3) \delta(\vk_4)}_c &= (2{\rm\pi})^3 \deltaD(\vk_{1234}) T(k_1,k_2,k_3,k_4) \ ,
\eal\eeq
where the subscript $c$ indicates the ``connected'' term.  Under our convention, $[P] = L^3$ and $[T] = L^6$.
For a given realization of the density field $\delta(\vk)$, the estimator of
the binned power spectrum is
\beq
\hat{P} (k_i) =\int_{k_i} \frac{{\rm d}^3 \vk}{V_s(k_i)}\delta(\vk)\delta(-\vk) \ ,
\eeq
where $V_s(k_i) = 4{\rm\pi} k_i^3 \delta\ln k$.
Its covariance is
\beq\bal
C_{ij} &= \avg{\hat P(k_i) \hat P(k_j)} - \avg{\hat P(k_i)} \avg{\hat
  P(k_j)} \label{eq:Cij} \\
&= \frac{(2{\rm\pi})^3}{V}\frac{2 P(k_i)^2}{V_s(k_i)} \delta_{ij} + \bar{T}(k_i, k_j)  \ ,
\eal\eeq
where 
\beq
\bar{T}(k_i, k_j) \equiv \int_{k_i}\frac{{\rm d}^3 \vk_1}{V_s(k_i)} \int_{k_j}\frac{{\rm d}^3 \vk_2}{V_s(k_j)} T(\vk_1,-\vk_1,\vk_2,-\vk_2)\ .  
\eeq
Below we provide the derivation.
The first term in equation (\ref{eq:Cij}) can be calculated as
\beq
\avg{\hat P(k_i) \hat P(k_j)} = \int_{k_i}\frac{{\rm d}^3 \vk_1}{V_s(k_i)} \int_{k_j}  
\frac{{\rm d}^3 \vk_2}{V_s(k_j)}
\avg{\delta(\vk_1)\delta(-\vk_1)\delta(\vk_2)\delta(-\vk_2)} \ ,
\eeq
where the integrand reads:
\beqa
&&\avg{\delta(\vk_1)\delta(-\vk_1)\delta(\vk_2)\delta(-\vk_2)} \nonumber\\
&&= \avg{\delta_1\delta_1^*\delta_2\delta_2^*}_c 
+\avg{\delta_1\delta_1^*} \avg{\delta_2\delta_2^*}
+\avg{\delta_1\delta_2} \avg{\delta_1^*\delta_2^*}
+\avg{\delta_1\delta_2^*} \avg{\delta_2\delta_1^*} \nonumber \\[0.2cm]
&&= (2{\rm\pi})^3 \deltaD(0) T(\vk_1,-\vk_1,\vk_2,-\vk_2) \label{eq:i}\\[0.2cm]
&&+ (2{\rm\pi})^6 \deltaD(0) P(\vk_1) \delta(0) P(\vk_2) \label{eq:ii}\\[0.2cm]
&&+ (2{\rm\pi})^6 \deltaD(\vk_1+\vk_2) P(\vk_1) \delta(\vk_1+\vk_2) P(\vk_1) \label{eq:iii}\\[0.2cm]
&&+ (2{\rm\pi})^6 \deltaD(\vk_1-\vk_2) P(\vk_1) \delta(\vk_1-\vk_2) P(\vk_1) \label{eq:iv}  \ .
\eeqa
We note that 
$\deltaD(0) = \frac{1}{(2{\rm\pi})^3}$ .
Then the contribution from each term reads
\beq\bal
\mbox{(\ref{eq:i})} &\Rightarrow \int_{k_i}\frac{{\rm d}^3 \vk_1}{V_s(k_i)} \int_{k_j}\frac{{\rm d}^3 \vk_2}{V_s(k_j)} T(\vk_1,-\vk_1,\vk_2,-\vk_2)
\equiv \bar{T}(k_i, k_j)  \\[0.2cm]
\mbox{(\ref{eq:ii})} &\Rightarrow \avg{\hat{P}(k_i)}\avg{\hat{P}(k_j)} \\
&\quad\mbox{(cancels the second term of equation (\ref{eq:Cij}))}\\[0.2cm]
\mbox{(\ref{eq:iii})} = \mbox{(\ref{eq:iv})} &\Rightarrow 
\int_{k_i}\frac{{\rm d}^3 \vk_1}{V_s(k_i)} \int_{k_j}\frac{{\rm d}^3 \vk_2}{V_s(k_j)} (2{\rm\pi})^3\deltaD(\vk_1-\vk_2) P(\vk_1) P(\vk_1) \\[0.2cm]
&= \int_{k_i}\frac{{\rm d}^3 \vk_1}{V_s(k_i)} P(k_1)^2 (2{\rm\pi})^3  \int_{k_j}\frac{{\rm d}^3 \vk_2}{V_s(k_j)} \deltaD(\vk_1-\vk_2) \\
&\quad\mbox{(only non-zero if $k_i = k_j$)} \\[0.2cm]
&= {\frac{(2{\rm\pi})^3}{V_z}\frac{\avg{\hat P(k_i)^2}}{V_s(k_i)} \delta_{ij} }
\approx \frac{(2{\rm\pi})^3}{V_z}\frac{P(k_i)^2}{V_s(k_i)} \delta_{ij} 
\eal\eeq
The expression of $\bar{T}(k_i, k_j)$ (equation \ref{eq:T1h}) can be obtained using equation~(\ref{eq:delta_g}) and 
is similar to the derivation of $P(k)$.

\end{document}